\documentclass[12pt,a4paper]{article}
\usepackage{amsmath,caption}
\usepackage[top=1.2in, bottom=1.2in, left=1.22in, right=1.22in]{geometry}

\DeclareCaptionStyle{italic}[justification=centering]{labelfont={bf},textfont={it},labelsep=colon}
\usepackage{graphicx,psfrag,epsf}
\usepackage{enumerate}
\usepackage{natbib}
\usepackage{url} 
\usepackage{booktabs, subfig, bm, paralist,mathpazo,tikz,todonotes,longtable,microtype,dsfont,rotating} 
\usepackage[pdftex,colorlinks=true]{hyperref}
\definecolor{darkblue}{rgb}{0,0,.6}
\hypersetup{citecolor=darkblue,linkcolor=darkblue,urlcolor=darkblue}

\newcommand{\blind}{0}

\graphicspath{{plots/}}
\DeclareMathOperator*{\argmin}{\arg\!\min}
\newsavebox\CBox
\def\textBF#1{\sbox\CBox{#1}\resizebox{\wd\CBox}{\ht\CBox}{\textbf{#1}}}

\definecolor{a0}{rgb}{0.0, 0.5, 0.0}
\definecolor{bistre}{rgb}{0.24, 0.17, 0.12}
\definecolor{amethyst}{rgb}{0.6, 0.4, 0.8}
\definecolor{blue-violet}{rgb}{0.54, 0.17, 0.89}
\definecolor{Rcolor}{RGB}{150,160,190}
\definecolor{blush}{rgb}{0.87, 0.36, 0.51}
\definecolor{brightturquoise}{rgb}{0.03, 0.91, 0.87}
\definecolor{burntorange}{rgb}{0.8, 0.33, 0.0}
\date{\today}

\begin{document}

\def\spacingset#1{\renewcommand{\baselinestretch}
{#1}\small\normalsize} \spacingset{1}

\if0\blind
{
  \title{\bf Grouped multivariate and functional time series forecasting: \hbox{an application to annuity pricing}}
  \author{Han Lin Shang\thanks{Postal address: Research School of Finance, Actuarial Studies and Statistics, Level 4, Building 26C, Australian National University, Kingsley Street, Canberra, ACT 2601, Australia; Telephone: +61(2) 612 50535; Fax: +61(2) 612 50087; Email: hanlin.shang@anu.edu.au.}
  \hspace{.2cm}\\
    Research School of Finance, Actuarial Studies and Statistics \\
    Australian National University \\
    and \\
    Steven Haberman \\
    CASS Business School, City University of London}
  \maketitle
} \fi

\if1\blind
{
  \bigskip
  \bigskip
  \bigskip
  \begin{center}
    {\LARGE\bf Title}
\end{center}
  \medskip
} \fi

\bigskip

\begin{abstract}

Age-specific mortality rates are often disaggregated by different attributes, such as sex, state, ethnic group and socioeconomic status. In making social policies and pricing annuity at national and subnational levels, it is important not only to forecast mortality accurately, but also to ensure that forecasts at the subnational level add up to the forecasts at the national level. This motivates recent developments in grouped functional time series methods \citep{SH16} to reconcile age-specific mortality forecasts. We extend these grouped functional time series forecasting methods to multivariate time series, and apply them to produce point forecasts of mortality rates at older ages, from which fixed-term annuities for different ages and maturities can be priced. Using the regional age-specific mortality rates in Japan obtained from the Japanese Mortality Database, we investigate the one-step-ahead to 15-step-ahead point-forecast accuracy between the independent and grouped forecasting methods. The grouped forecasting methods are shown not only to be useful for reconciling forecasts of age-specific mortality rates at national and subnational levels, but they are also shown to allow improved forecast accuracy. The improved forecast accuracy of mortality rates is of great interest to the insurance and pension industries for estimating annuity prices, in particular at the level of population subgroups, defined by key factors such as sex, region, and socioeconomic grouping. 

\end{abstract}

\\

\newpage
\spacingset{1.48}

\section{Introduction}

In many developed countries such as Japan, increases in longevity and an aging population have led to concerns about the sustainability of pensions, healthcare and aged-care systems \citep[e.g.,][]{Coulmas07, OECD13}. These concerns have resulted in a surge of interest among government policy makers and planners to engage in accurate modeling and forecasting of age-specific mortality rates. Any improvements in the forecast accuracy of mortality would be beneficial for annuity providers, corporate pension funds and governments \citep[e.g.,][]{Koissia06, DDG07, HPG11}, in particular for determining age of retirement and allocating pension benefits at the national and subnational levels.

Several authors have proposed new approaches for forecasting age-specific mortality at the national level using statistical models \citep[for reviews, see][]{Booth06, BT08}. These models can be categorized by the approach used into three main streams: explanation, expectation, and extrapolation approaches. \citeauthor{LC92}'s \citeyearpar{LC92} work represents a significant milestone in demographic forecasting employing the extrapolation method, and the so-called `Lee--Carter model' has since received considerable attention in demography and actuarial science. This model has been extensively studied and widely used for forecasting mortality rates in various countries \citep[see][and the references therein]{SBH11}. 

The strengths of the Lee--Carter method are its simplicity and robustness in situations where age-specific log mortality rates have linear trends \citep{BHT+06}. The main weakness of the Lee--Carter method is that it attempts to capture the patterns of mortality rates using only one principal component and its scores. To rectify this deficiency, the Lee--Carter model has been extended and modified in several directions \citep[e.g.,][]{BDV02, RH03, CDE04, RH06, HU07, PDH+09}.

Although mortality forecasts at the national level are comparably accurate, mortality forecasts at the subnational level often suffer from relatively poor data quality and/or missing data. However, subnational forecasts of age-specific mortality are valuable for informing policy within local regions, and allow the appreciation of the heterogeneity in the population and the understanding of differences between individual. A better understanding of individual characteristics allows assurers to better price annuity products for annuitants. 

In insurance and pension companies, it is typically of interest to forecast age-specific mortality for multiple subpopulations that often obey a hierarchical (unique) or group (non-unique) structure. Let us consider a simple group structure, where total age-specific mortality rates can be disaggregated by sex. If female, male and total age-specific mortality are forecasted independently, the forecast female and male mortality may not add up to the forecast total mortality. This is known as the problem of forecast reconciliation, which bears a strong resemblance to the issue of forecasting contemporal time series aggregates \citep[e.g.,][]{WA81,Lutkepohl84, HH11}. Similar to contemporal time series aggregation, among our grouped time series forecasting methods, we also consider the bottom-up method where the forecasts at the aggregated levels can be obtained by forecasting the most disaggregated series. However, differing from these early works, each series in our context is a time series of functions instead of a univariate time series.

Forecast reconciliation has been considered in economics for balancing national accounts \citep[e.g.,][]{SCM42}, for forecasting aggregate United States  inflation \citep[e.g.,][]{HH11}, and for forecasting personal consumption expenditures \citep[e.g.,][]{Lutkepohl84}. In addition, forecast reconciliation has been considered in statistics for forecasting tourism demand \citep{HAA+11}, in operation research for forecasting demand for accident and emergency services in the United Kingdom \citep{AHK+17}, and in demography for forecasting age-specific mortality rates \citep{Shang16, SH16}. To the best of our knowledge, forecast reconciliation of age-specific mortality has not been considered in actuarial studies to date and it is our goal to fill this methodological gap. 

We apply two forecasting techniques, the Lee--Carter method and the functional time series method of \cite{HU07}, to a large set of multivariate or functional time series with rich structure, respectively. We put forward two statistical methods, the bottom-up and optimal-combination methods, to reconcile point forecasts of age-specific mortality, and potentially improve the point-forecast accuracy. This approach may lead to more accurate forecasts of mortality and conditional life expectancy, thus better estimates of annuity prices. The bottom-up method involves forecasting each of the disaggregated series and then using simple aggregation to obtain forecasts for the aggregated series \citep{Kahn98}. This method works well when the bottom-level series have high signal-to-noise ratio. For highly disaggregated series, this does not work well because the series become too noisy. This motivates the development of the optimal-combination method \citep{HAA+11}, where forecasts are obtained independently for all series at all levels of disaggregation and then a linear regression is used with an ordinary least-squares or a generalized least-squares estimator to optimally combine and reconcile these forecasts.

Using the national and subnational Japanese age-specific mortality rates from 1975 to 2014, we compare the point-forecast accuracy among the independent (base) forecasting, bottom-up and optimal-combination methods. The independent forecasts can be produced from a multivariate or functional time series forecasting method. These independent forecasts are generally not reconciled according to the group structure. To evaluate point-forecast accuracy, we consider the mean absolute forecast error (MAFE) and the root mean square forecast error (RMSFE), and find that the bottom-up method performs the best among the three methods in our data set. 

The remainder of this paper is structured as follows: In Section~\ref{sec:2}, we describe the motivating data set, which is Japanese national and subnational age-specific mortality rates. In Section~\ref{sec:3}, we briefly revisit the Lee--Carter and functional time series methods for producing point forecasts. In Section~\ref{sec:4}, we introduce two grouped forecasting methods . Using the forecast-error criteria in Section~\ref{sec:point_eval}, we first evaluate and compare point-forecast accuracy between the Lee--Carter and the functional time series methods, and then between the independent and grouped forecasting methods in Section~\ref{sec:point_compar}. In Section~\ref{sec:interval_section}, we provide results for the interval forecasts. In Section~\ref{sec:annuity}, we apply the independent and grouped forecasting methods to estimate the fixed-term annuity prices for different ages and maturities. Conclusions are presented in Section~\ref{sec:conclu}, along with some reflections on how the methods presented here can be extended.

\section{Data}\label{sec:2}

We study Japanese age-specific mortality rates from 1975 to 2014, obtained from the Japanese Mortality Database \citep{JMD15}. Given that our focus is on life-annuities pricing, we consider ages from 60 to 99 in a single year of age, and the last age group is the age at and beyond 100. The structure of the data is presented in Table~\ref{tab:1} where each row denotes a level of disaggregation. 

\begin{table}[!htbp]
\centering
\tabcolsep 0.32in
\caption{Hierarchy of Japanese mortality rates}\label{tab:1}
\begin{tabular}{@{}lr@{}}
\toprule
Group level &  Number of series \\\midrule
Japan & 1 \\
Sex & 2 \\
Region & 8 \\
Region $\times$ Sex & 16 \\
Prefecture & 47 \\
Prefecture $\times$ Sex & 94 \\\midrule
Total & 168 \\\bottomrule
\end{tabular}
\end{table}
At the top level, we have total age-specific mortality rates for Japan. We can split these total mortality rates by sex, region or prefecture. There are eight regions in Japan, which contain a total of 47 prefectures. The most disaggregated data arise when we consider the mortality rates for each combination of prefecture and sex, giving a total of $47\times 2 = 94$ series. In total, across all levels of disaggregation, there are 168 series.  

\subsection{Rainbow plots}

\begin{figure}[!htbp]
\centering
\subfloat[Observed female mortality rates]
{\includegraphics[width=8.2cm]{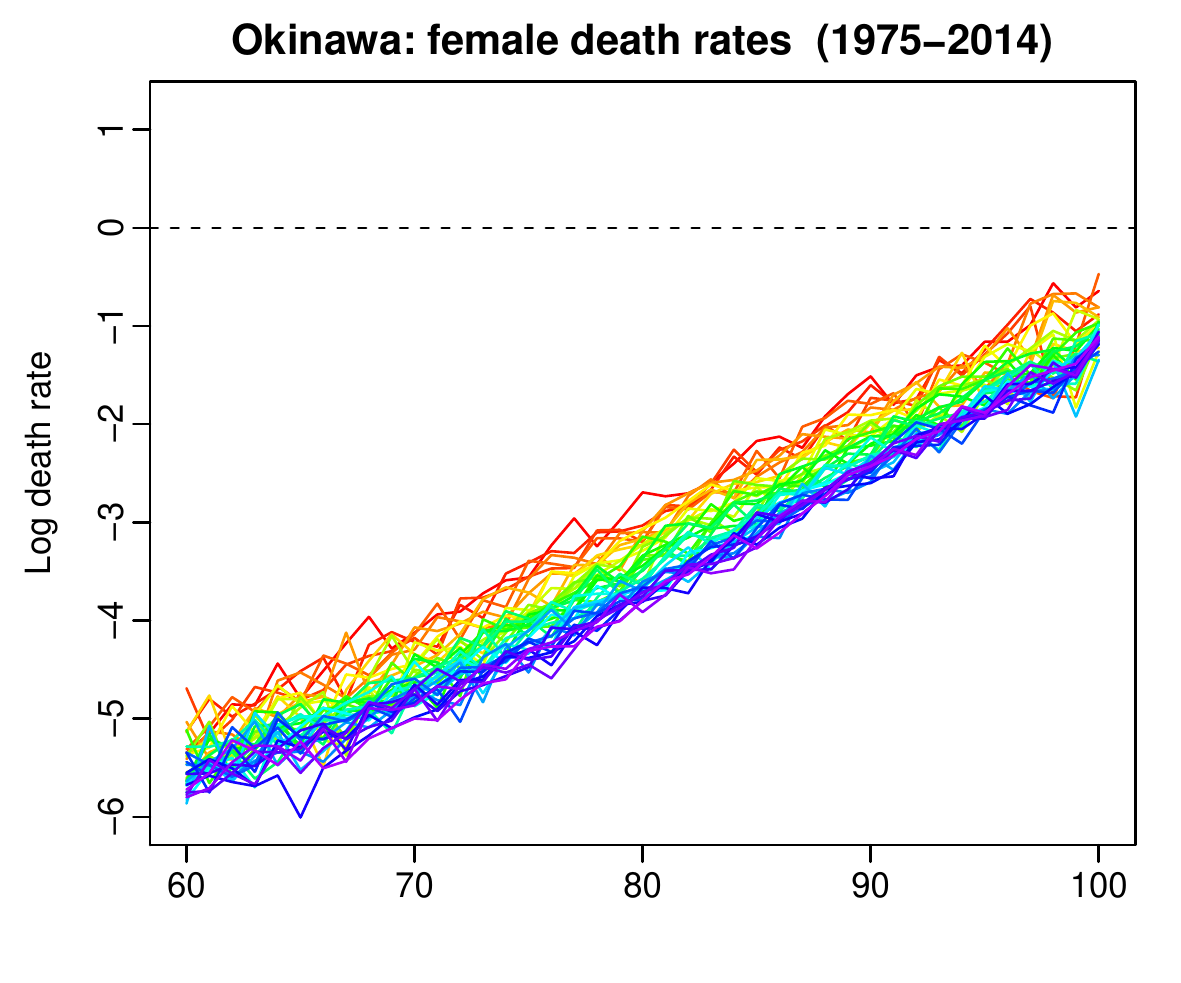}\label{fig:1a}}
\quad
\subfloat[Observed male mortality rates]
{\includegraphics[width=8.2cm]{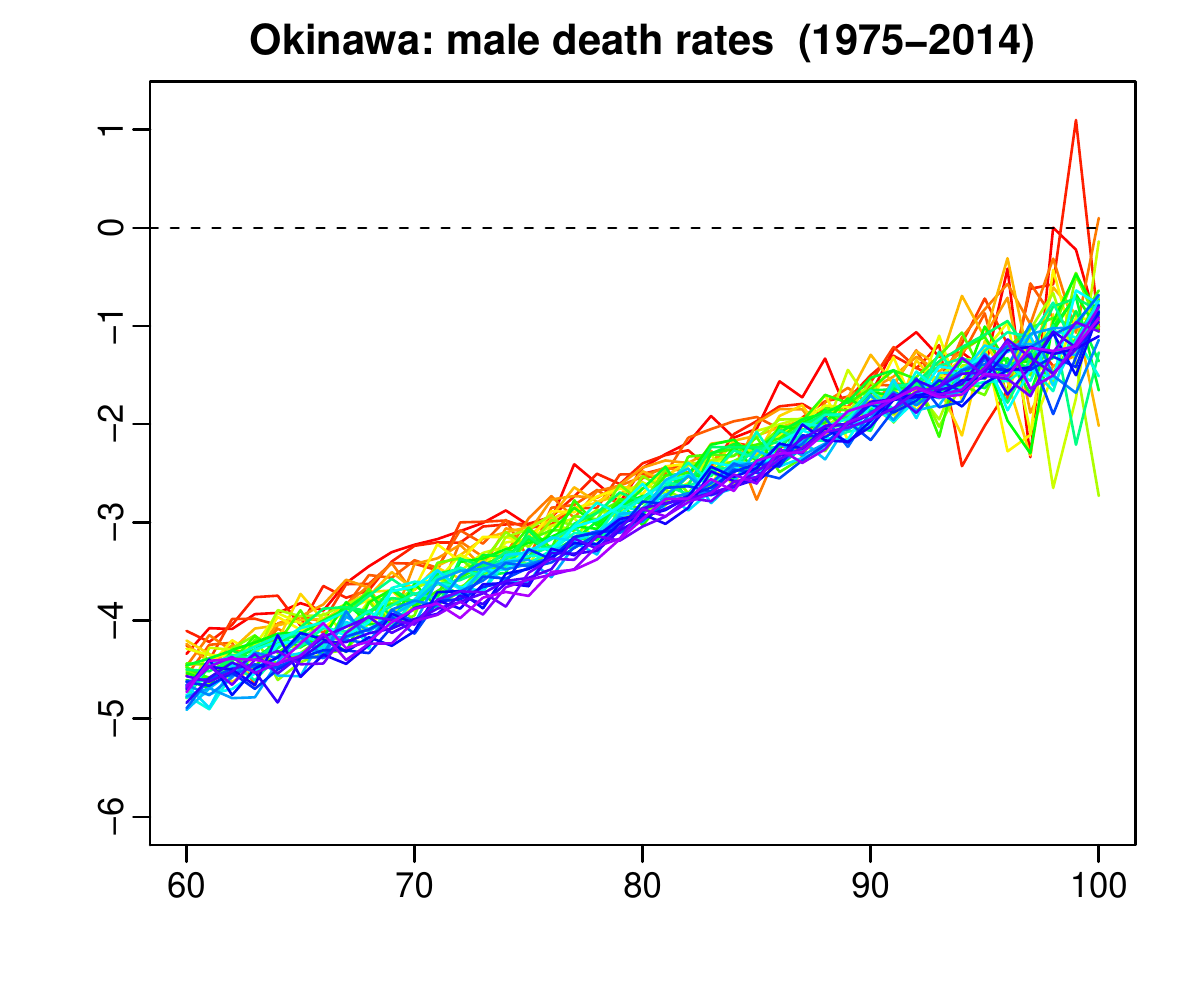}\label{fig:1b}}
\\
\subfloat[Smoothed female mortality rates]
{\includegraphics[width=8.2cm]{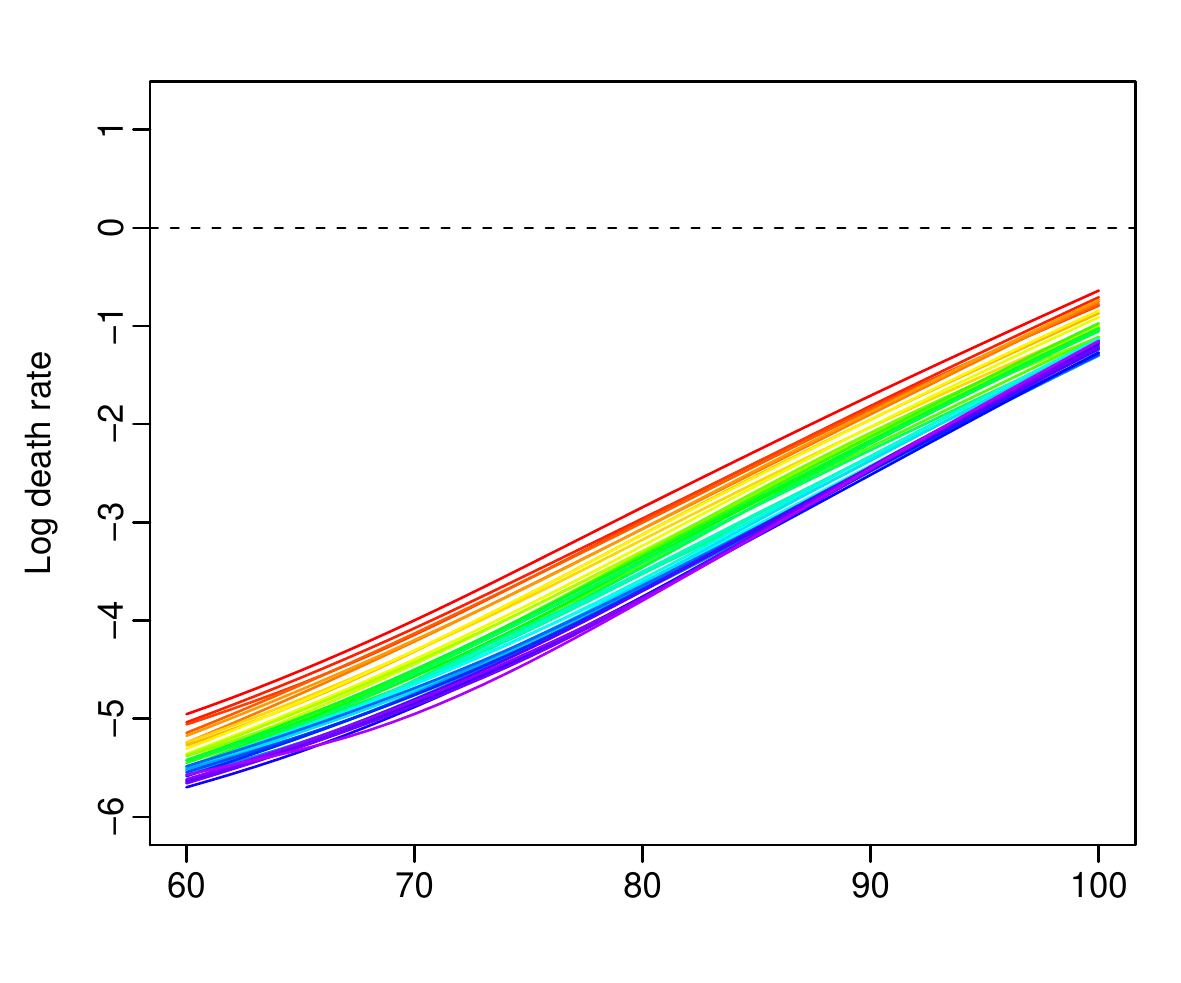}\label{fig:1c}}
\quad
\subfloat[Smoothed male mortality rates]
{\includegraphics[width=8.2cm]{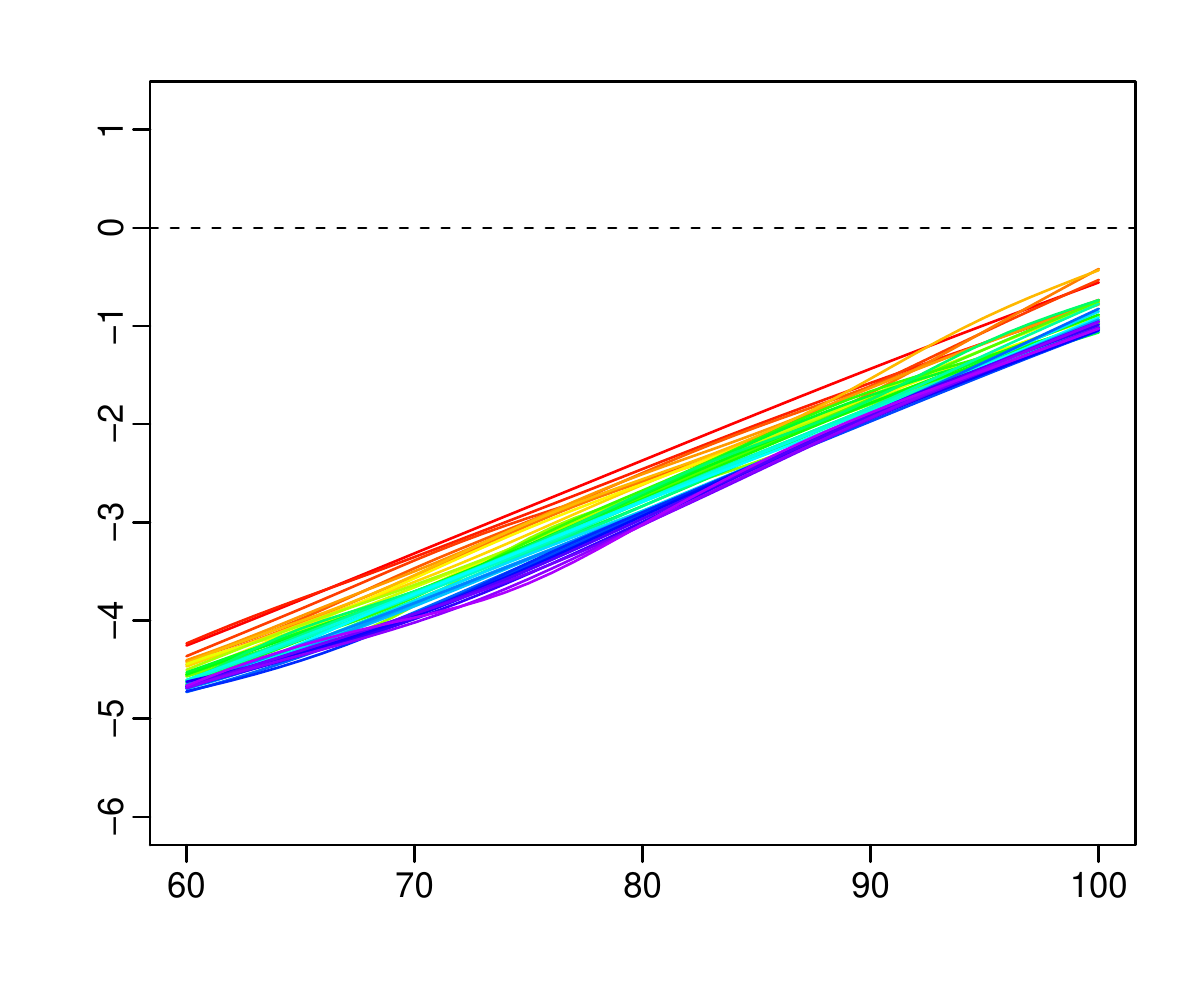}\label{fig:1d}}
\caption{Functional time series graphical displays.}\label{fig:1}
\end{figure}

Figure~\ref{fig:1} presents rainbow plots of the female and male age-specific log mortality rates in the prefecture of Okinawa from 1975 to 2014. The time ordering of the curves follows the color order of a rainbow, where curves from the distant past are shown in red and the more recent curves are shown in purple \citep{HS10}. The figures demonstrate typical age-specific mortality curves with gradually increasing mortality rates as age increases.

Figures~\ref{fig:1a} and~\ref{fig:1b} demonstrate that the observed mortality rates are not smooth across age. Due to observational noise, male mortality rates in some years are above 1 (when log mortality rates are above 0). To obtain smooth functions and deal with possible missing values, we consider a penalized regression spline smoothing with monotonic constraint, described in Section~\ref{sec:32}. The penalized regression spline smoothing with monotonic constraint incorporates the shape of log mortality curves \citep[see also][]{HU07, DPR11}.

Figures~\ref{fig:1c} and~\ref{fig:1d} present the smooth age-specific mortality rates for Okinawa females and males, but we apply smoothing to all series at different levels of disaggregation. We developed a Shiny application \citep{Chang16} in R \citep{Team16} to allow interactive exploration of the smoothing of all the data series, which is available in the online supplement. 

\subsection{Image plots}

\begin{figure}[!htbp]
\centering
\includegraphics[width=15.8cm]{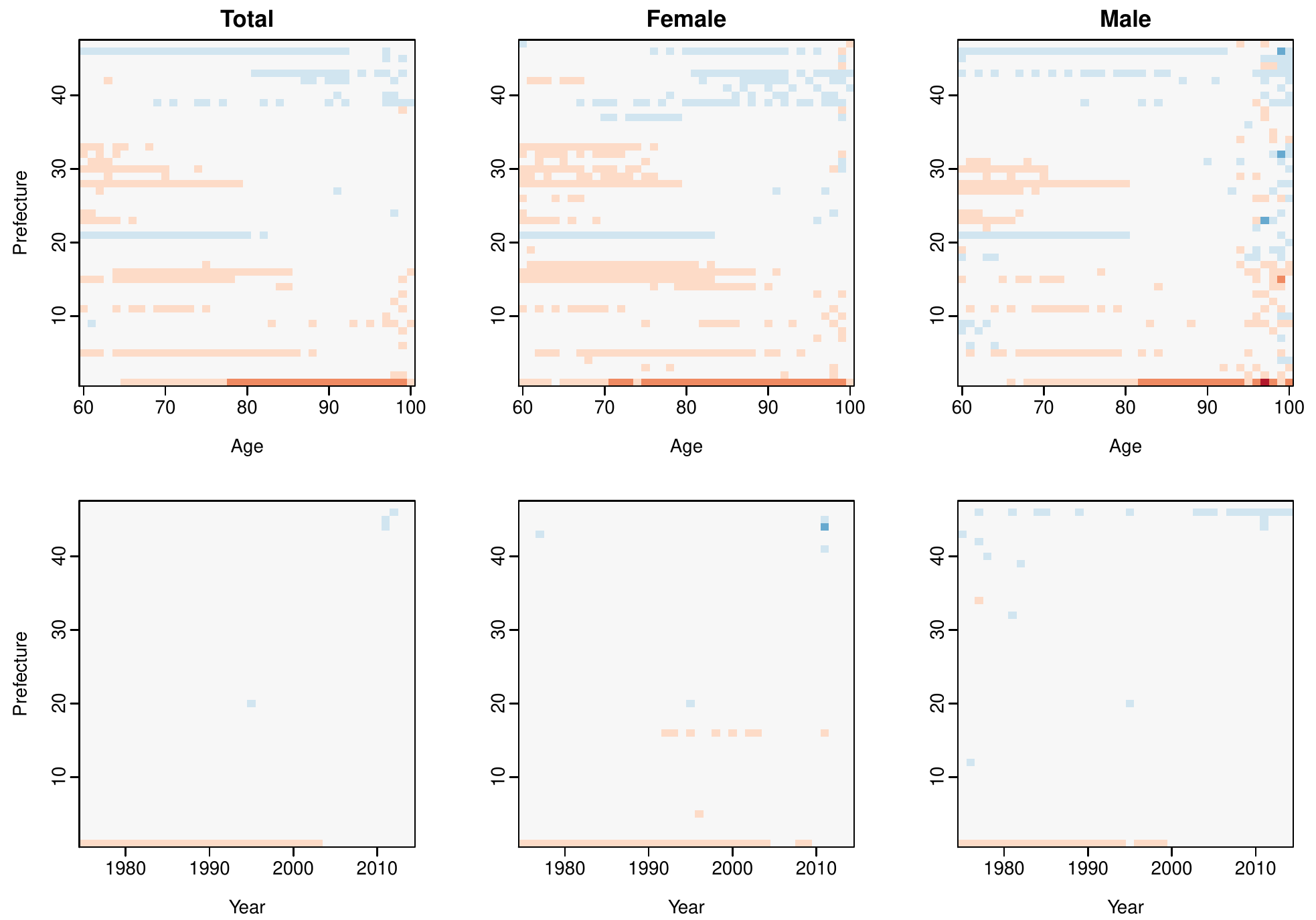}
\caption{Image plots showing log of the ratios of mortality rates. The top panel shows mortality rates averaged over years, while the bottom panel shows mortality rates averaged over ages.}\label{fig:image}
\end{figure}

Another visual perspective of the data is the image plot of \cite{SH16}. In Figure~\ref{fig:image}, we plot the log of the ratio of mortality rates for each prefecture to the mortality rates for Japan, because this facilitates relative mortality comparison. A divergent color palette is used with blue representing positive values and orange representing negative values. The prefectures are ordered geographically from north (Hokkaido) to south (Okinawa). 

The top row of panels shows mortality rates for each prefecture and age, averaged over years. There are strong differences between the prefectures for the elderly; this is possibly due to differences in socioeconomic status and accessibility of health services. The most southerly prefecture of Okinawa has very low mortality rates and thus extreme longevity for the elderly \citep[e.g.,][]{takata1987influence, suzuki2004successful,willcox2007aging}. 

The bottom row of panels shows mortality rates for each prefecture and year, averaged over all ages. We find three abnormalities. In 2011, in prefectures 44 (Miyagi) and 45 (Iwate), there are abnormally large increases in mortality compared to the other prefectures. These are northern coastal regions, and the inflated relative mortality rates are due to the tsunami that occurred on 11 March 2011 \citep{SH16}. In 1995, there is an abnormal increase in mortality for prefecture 20 (Hy$\bar{\text{o}}$go), which corresponds to the Kobe (Great Hanshin) earthquake of 17 January 1995. In the prefecture Okinawa, the residents enjoy relatively low mortality rates until 2000 and even beyond, particularly for females. However, recently, the comparably lower mortality rates become less evident. 

\section{Forecasting methods}\label{sec:3}

We revisit the Lee--Carter and functional time series methods for forecasting age-specific mortality, and the forecast accuracy of both methods is compared in the present study. The Lee--Carter model considers age a discrete variable \citep[e.g.,][]{Bell92, LL05}, while the functional time series model treats age as a continuous variable \citep[e.g.,][]{DPR11,Shang16b}. To stabilize the high variance associated with high age-specific mortality rates, it is necessary to transform the raw data by taking the natural logarithm. We denote by $m_{x,t}$ the observed mortality rate at age $x$ in year $t$ calculated as the number of deaths aged $x$ in calendar year $t$, divided by the corresponding mid-year population aged $x$. The models are all expressed on the log scale.

\subsection{Lee--Carter method}\label{sec:31}

The original formulation of the Lee--Carter model is given by
\begin{equation}
\ln (m_{x,t}) = a_x + b_x \kappa_t + \varepsilon_{x,t}, \label{eq:1}
\end{equation}
where $a_x$ is the age pattern of the log mortality rates averaged across years; $b_x$ is the first principal component reflecting relative change in the log mortality rate at each age; $\kappa_t$ is the first set of principal component scores at year $t$ and measures the general level of the log mortality rates; and $\varepsilon_{x,t}$ is the residual at age $x$ and year $t$.

The Lee--Carter model in~\eqref{eq:1} is over-parametrized in that the model structure is invariant under the following transformations:
\begin{align*}
\{a_x, b_x, \kappa_t\}  &\mapsto \{a_x, b_x/c, c\kappa_t\}, \\
\{a_x, b_x, \kappa_t\}  &\mapsto \{a_x - c b_x, b_x, \kappa_t + c\}
\end{align*}
To ensure the model identifiability, \cite{LC92} imposed two constraints given as
\begin{equation*}
\sum^n_{t=1}\kappa_t = 0, \qquad \sum_{x=x_1}^{x_p}b_x = 1,
\end{equation*}
where $n$ denotes the number of years and $p$ denotes the number of ages in the observed data set.

The Lee--Carter method adjusts $\kappa_t$ by refitting the total number of deaths. The adjustment gives more weight to high rates \citep{SBH11}. The adjusted $\kappa_t$ is then extrapolated using autoregressive integrated moving average (ARIMA) models. \cite{LC92} used a random walk with drift model, which can be expressed as
\begin{equation*}
\kappa_t = \kappa_{t-1} + d + e_t,
\end{equation*}
where $d$ is known as the drift parameter and measures the average annual change in the series, and $e_t$ is an uncorrelated error. Based on the forecast of principal component scores, the forecast age-specific log mortality rates are obtained using the estimated mean function $\widehat{a}_x$ and estimated first principal component $\widehat{b}_x$ in~\eqref{eq:1}.

\subsection{A functional time series method}\label{sec:32}

The Lee--Carter model considers age a discrete variable, while the functional time series model treats age as a continuous variable. One advantage of the functional time series model is that a nonparametric smoothing technique can be incorporated into the modeling procedure to obtain smoothed principal components. Smoothing deals with the criticism of the Lee--Carter model that the estimated values, $b_x$, can be subject to considerable noise and, without smoothing, this would be incorporated into forecasts of future mortality rates.

Among the many possible nonparametric smoothing techniques, we use penalized regression spline with a partial monotonic constraint, where the smoothed log mortality rates can be expressed as
\begin{equation*}
m_t(x_i) = f_t(x_i) + \sigma_t(x_i)\varepsilon_{t,i}, \qquad i=1,\dots,p, \quad t=1,\dots,n,
\end{equation*}
where $m_t(x_i)$ denotes the log of the observed mortality rate for age $x_i$ in year $t$; $\sigma_t(x_i)$ allows the amount of noise to vary with $x_i$ in year $t$; and $\varepsilon_{t,i}$ is an independent and identically distributed standard normal random variable. 

The smoothed log mortality curves $\bm{f}(x) = \{f_1(x),\dots,f_n(x)\}$ are treated as realizations of a stochastic process. Using functional principal component analysis, these smoothed log mortality curves are decomposed into 
\begin{equation}
f_t(x) = a(x) + \sum^J_{j=1}b_j(x)k_{t,j} + e_t(x), \label{eq:decomp}
\end{equation}
where $a(x)$ denotes the mean function, $\{b_1(x),\dots,b_J(x)\}$ denotes a set of functional principal components, $\{k_{t,1},\dots,k_{t,J}\}$ denotes a set of principal component scores in year $t$, $e_t(x)$ is the error function with mean zero, and $J<n$ is the number of principal components retained. Decomposition~\eqref{eq:decomp} facilitates dimension reduction because the first $J$ terms often provide a reasonable approximation to the infinite sums, and thus the information contained in $\bm{f}(x)$ can be adequately summarized by the $J$-dimensional vector $(\bm{b}_1,\dots,\bm{b}_J)$. In contrast to the Lee--Carter model, another advantage of the functional time series model is that more than one component may be used to improve model fitting \citep[see also][]{RH03}.

Conditioning on the observed data $\bm{\mathcal{I}} = \{m_1(x),\dots,m_n(x)\}$ and the set of functional principal components $\bm{B} = \{b_1(x),\dots,b_J(x)\}$, the $h$-step-ahead forecast of $m_{n+h}(x)$ can be obtained by
\begin{align*}
\widehat{m}_{n+h|n}(x) &= \text{E}[m_{n+h}(x)|\bm{\mathcal{I}},\bm{B}] \\
&= \widehat{a}(x) + \sum^J_{j=1}b_j(x)\widehat{k}_{n+h|n,j},
\end{align*}
where $\widehat{k}_{n+h|n,j}$ denotes the $h$-step-ahead forecast of $k_{n+h,j}$ using a univariate or multivariate time series model \citep[for further detail, see][]{HS09, ANH15}. Here, we consider a univariate time series forecasting method and implement the automatic algorithm of \cite{HK08} for selecting optimal orders in the ARIMA model. Having identified the optimal ARIMA model, the maximum likelihood method can be used to
estimate the parameters.

To select $J$, we determine the value of $J$ as the minimum number of components that reaches a certain level of the proportion of total variance explained by the $J$ leading components such that
\begin{equation*}
J = \argmin_{J: J\geq 1}\left\{\sum^J_{j=1}\widehat{\lambda}_j\Bigg/\sum^{\infty}_{j=1}\widehat{\lambda}_j\mathds{1}\{\widehat{\lambda}_j>0\}\geq \delta\right\},
\end{equation*}
where $\delta = 95\%$, and $\mathds{1}\{\cdot\}$ denotes the binary indicator function which excludes possible zero eigenvalues. For all the series, the first functional principal component can explain at least 90\% of total variation. As a sensitivity test, we also consider $\delta=99\%$, as well as $J=6$ \citep[see][]{HBY13}. In the online supplement, we also present the results of the point and interval forecasts for different values of $J$ determined by the two different criteria.

\section{Grouped forecasting methods}\label{sec:4}

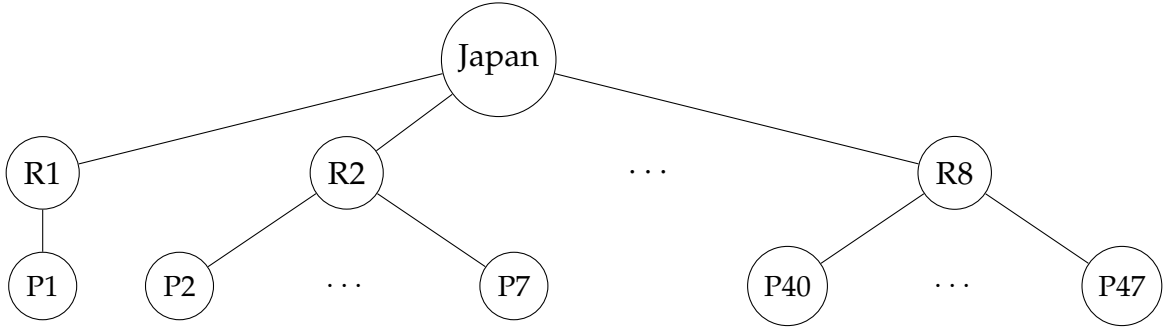
\begin{figure}[!htb]
\centering\begin{tikzpicture}
\tikzstyle{every node}=[minimum size = 8mm]
\tikzstyle[level distance=10cm] \tikzstyle[sibling distance=40cm]
\tikzstyle{level 3}=[sibling distance=16mm,font=\footnotesize]
\tikzstyle{level 2}=[sibling distance=22mm,font=\small]
\tikzstyle{level 1}=[sibling distance=40mm,font=\normalsize]
\node[circle,draw]{Japan}
   child {node[circle,draw] {R1}
   	     child {node[circle,draw] {P1}}}
   child {node[circle,draw] {R2}
   		child {node[circle,draw] {P2}}
      child {node {$\cdots$}edge from parent[draw=none]}
		child {node[circle,draw] {P7}}
		}
  child {node {$\cdots$}edge from parent[draw=none]}
   child {node[circle,draw] {R8}
   		child{node[circle,draw] {P40}}
	     child{node {$\cdots$}edge from parent[draw=none]}
  	child{node[circle,draw] {P47}}
 };
\end{tikzpicture}
\medskip
\caption{The Japanese geographical hierarchy tree diagram, with eight regions and 47 prefectures--each node has female, male and total age-specific mortality rates.}\label{fig:2}
\end{figure}

For ease of explanation, we will introduce the grouped forecasting methods using the Japanese example provided in Section~\ref{sec:2}. The Japanese data follow a three-level geographical hierarchy, coupled with a sex-grouping variable. The geographical hierarchy is presented in Figure~\ref{fig:2}. Japan can be split into eight regions from north to south, which are then split into 47 prefectures.

\medskip

The data can also be split by sex. Each of the nodes in the geographical hierarchy can also be split into both males and females. We refer to a particular disaggregated series using the notation $\text{X} \ast \text{S}$, referring to the geographical area $\text{X}$ and the sex $\text{S}$, where $\text{X}$ can take the values shown in Figure~\ref{fig:2} and $\text{S}$ can take values M (males), F (females) or T (total). For example, $\text{R1}\ast \text{F}$ denotes females in Region 1; $\text{P1}\ast \text{T}$ denotes all females and males in Prefecture 1; $\text{Japan}\ast \text{M}$ denotes all males in Japan.

Denote E$_{\text{X}\ast \text{S}, t}(x)$ as the exposure-to-risk for series $\text{X}\ast \text{S}$ in year $t$ and age $x$, and let $\text{D}_{\text{X}\ast \text{S}, t}(x)$ be the number of deaths for series $\text{X}\ast \text{S}$ in year $t$ and age $x$. The age-specific mortality rate is then given by 
\begin{equation*}
\text{R}_{\text{X}\ast \text{S}, t}(x) = \text{D}_{\text{X}\ast \text{S}, t}(x)\big/\text{E}_{\text{X}\ast \text{S}, t}(x).
\end{equation*}
To simplify expressions, we will drop the age argument $(x)$. Then for a given age, we can write
\[
\hspace{-.2in} \underbrace{ \left[
\begin{footnotesize}
\begin{array}{l}
\text{R}_{\text{Japan*T},t} \\
\text{R}_{\textcolor{red}{\text{Japan*F},t}} \\
\text{R}_{\textcolor{red}{\text{Japan*M},t}} \\
\text{R}_{\textcolor{a0}{\text{R1*T},t}} \\
\text{R}_{\textcolor{a0}{\text{R2*T},t}} \\
\vdots \\
\text{R}_{\textcolor{a0}{\text{R8*T},t}} \\
\text{R}_{\textcolor{blue-violet}{\text{R1*F},t}} \\
\text{R}_{\textcolor{blue-violet}{\text{R2*F},t}} \\
\vdots \\
\text{R}_{\textcolor{blue-violet}{\text{R8*F},t}} \\
\text{R}_{\textcolor{burntorange}{\text{R1*M},t}} \\
\text{R}_{\textcolor{burntorange}{\text{R2*M},t}} \\
\vdots \\
\text{R}_{\textcolor{burntorange}{\text{R8*M},t}} \\
\text{R}_{\textcolor{blue}{\text{P1*T},t}} \\
\text{R}_{\textcolor{blue}{\text{P2*T},t}} \\
\vdots \\
\text{R}_{\textcolor{blue}{\text{P47*T},t}} \\
\text{R}_{\textcolor{purple}{\text{P1*F},t}} \\
\text{R}_{\textcolor{purple}{\text{P1*M},t}} \\
\text{R}_{\textcolor{purple}{\text{P2*F},t}} \\
\text{R}_{\textcolor{purple}{\text{P2*M},t}} \\
\vdots \\
\text{R}_{\textcolor{purple}{\text{P47*F},t}} \\
\text{R}_{\textcolor{purple}{\text{P47*M},t}} \\ \end{array}
\end{footnotesize} \right]}_{\bm{R}_t} =
\underbrace{\left[
\begin{footnotesize}
\begin{array}{ccccccccccc}
\frac{\text{E}_{\text{P1*F},t}}{\text{E}_{\text{Japan*T},t}} & \frac{\text{E}_{\text{P1*M},t}}{\text{E}_{\text{Japan*T},t}} & \frac{\text{E}_{\text{P2*F},t}}{\text{E}_{\text{Japan*T},t}} & \frac{\text{E}_{\text{P2*M},t}}{\text{E}_{\text{Japan*T},t}}  & \frac{\text{E}_{\text{P3*F},t}}{\text{E}_{\text{Japan*T},t}} & \frac{\text{E}_{\text{P3*M},t}}{\text{E}_{\text{Japan*T},t}} & \cdots & \frac{\text{E}_{\text{P47*F},t}}{\text{E}_{\text{Japan*T},t}} & \frac{\text{E}_{\text{P47*M},t}}{\text{E}_{\text{Japan*T},t}} \\
\textcolor{red}{\frac{\text{E}_{\text{P1*F},t}}{\text{E}_{\text{Japan*F},t}}} & \textcolor{red}{0} & \textcolor{red}{\frac{\text{E}_{\text{P2*F},t}}{\text{E}_{\text{Japan*F},t}}} & \textcolor{red}{0} & \textcolor{red}{\frac{\text{E}_{\text{P3*F},t}}{\text{E}_{\text{Japan*F},t}}} & \textcolor{red}{0} & \cdots & \textcolor{red}{\frac{\text{E}_{\text{P47*F},t}}{\text{E}_{\text{Japan*F},t}}} & \textcolor{red}{0} \\
\textcolor{red}{0} & \textcolor{red}{\frac{\text{E}_{\text{P1*M},t}}{\text{E}_{\text{Japan*M},t}}}  & \textcolor{red}{0} & \textcolor{red}{\frac{\text{E}_{\text{P2*M},t}}{\text{E}_{\text{Japan*M},t}}} & \textcolor{red}{0} & \textcolor{red}{\frac{\text{E}_{\text{P3*M},t}}{\text{E}_{\text{Japan*M},t}}} & \cdots & \textcolor{red}{0} & \textcolor{red}{\frac{\text{E}_{\text{P47*M},t}}{\text{E}_{\text{Japan*M},t}}} \\
\textcolor{a0}{\frac{\text{E}_{\text{P1*F},t}}{\text{E}_{\text{R1,T},t}}} & \textcolor{a0}{\frac{\text{E}_{\text{P1*M},t}}{\text{E}_{\text{R1,T},t}}} & \textcolor{a0}{0} & \textcolor{a0}{0} & \textcolor{a0}{0} & \textcolor{a0}{0} & \cdots  & \textcolor{a0}{0} & \textcolor{a0}{0} \\
\textcolor{a0}{0} & \textcolor{a0}{0} & \textcolor{a0}{\frac{\text{E}_{\text{P2*F},t}}{\text{E}_{\text{R2,T},t}}} & \textcolor{a0}{\frac{\text{E}_{\text{P2*M},t}}{\text{E}_{\text{R2,T},t}}} & \textcolor{a0}{\frac{\text{E}_{\text{P3*F},t}}{\text{E}_{\text{R2,T},t}}} & \textcolor{a0}{\frac{\text{E}_{\text{P3*M},t}}{\text{E}_{\text{R2,T},t}}} & \cdots & \textcolor{a0}{0} & \textcolor{a0}{0} \\
\vdots & \vdots & \vdots & \vdots & \vdots & \vdots & \cdots & \vdots & \vdots \\
\textcolor{a0}{0} & \textcolor{a0}{0} & \textcolor{a0}{0} & \textcolor{a0}{0} & \textcolor{a0}{0} & \textcolor{a0}{0} & \cdots & \textcolor{a0}{\frac{\text{E}_{\text{P47*F},t}}{\text{E}_{\text{R8,T},t}}} & \textcolor{a0}{\frac{\text{E}_{\text{P47*M},t}}{\text{E}_{\text{R8,T},t}}} \\
\textcolor{blue-violet}{\frac{\text{E}_{\text{P1*F},t}}{\text{E}_{\text{R1,F},t}}} & \textcolor{blue-violet}{0} & \textcolor{blue-violet}{0} & \textcolor{blue-violet}{0} & \textcolor{blue-violet}{0} & \textcolor{blue-violet}{0} &  \cdots & \textcolor{blue-violet}{0} & \textcolor{blue-violet}{0} \\
\textcolor{blue-violet}{0} & \textcolor{blue-violet}{0} & \textcolor{blue-violet}{\frac{\text{E}_{\text{P2*F},t}}{\text{E}_{\text{R2,F},t}}} & \textcolor{blue-violet}{0} & \textcolor{blue-violet}{\frac{\text{E}_{\text{P3*F},t}}{\text{E}_{\text{R2,F},t}}} & \textcolor{blue-violet}{0} & \cdots & \textcolor{blue-violet}{0} & \textcolor{blue-violet}{0}  \\
\vdots & \vdots & \vdots & \vdots & \vdots & \vdots & \cdots & \vdots & \vdots \\
\textcolor{blue-violet}{0} & \textcolor{blue-violet}{0}  & \textcolor{blue-violet}{0}  & \textcolor{blue-violet}{0}  & \textcolor{blue-violet}{0}  & \textcolor{blue-violet}{0}  & \cdots & \textcolor{blue-violet}{\frac{\text{E}_{\text{P47*F},t}}{\text{E}_{\text{R8,F},t}}} & \textcolor{blue-violet}{0}\\
\textcolor{burntorange}{0} & \textcolor{burntorange}{\frac{\text{E}_{\text{P1*M},t}}{\text{E}_{\text{R1,M},t}}} & \textcolor{burntorange}{0} &\textcolor{burntorange}{0} & \textcolor{burntorange}{0} & \textcolor{burntorange}{0} & \cdots & \textcolor{burntorange}{0} & \textcolor{burntorange}{0} \\
\textcolor{burntorange}{0} & \textcolor{burntorange}{0} & \textcolor{burntorange}{0} & \textcolor{burntorange}{\frac{\text{E}_{\text{P2*M},t}}{\text{E}_{\text{R2,M},t}}} & \textcolor{burntorange}{0} & \textcolor{burntorange}{\frac{\text{E}_{\text{P3*M},t}}{\text{E}_{\text{R2,M},t}}} & \cdots & \textcolor{burntorange}{0} & \textcolor{burntorange}{0} \\
\vdots & \vdots & \vdots & \vdots & \vdots & \vdots & \cdots & \vdots & \vdots \\
\textcolor{burntorange}{0} & \textcolor{burntorange}{0} & \textcolor{burntorange}{0} & \textcolor{burntorange}{0} & \textcolor{burntorange}{0} & \textcolor{burntorange}{0} & \cdots & \textcolor{burntorange}{0} & \textcolor{burntorange}{\frac{\text{E}_{\text{P47*M},t}}{\text{E}_{\text{R8,M},t}}} \\
\textcolor{blue}{\frac{\text{E}_{\text{P1*F},t}}{\text{E}_{\text{P1,T},t}}} & \textcolor{blue}{\frac{\text{E}_{\text{P1*M},t}}{\text{E}_{\text{P1,T},t}}} & \textcolor{blue}{0} & \textcolor{blue}{0} & \textcolor{blue}{0} & \textcolor{blue}{0} & \cdots & \textcolor{blue}{0} & \textcolor{blue}{0} \\
\textcolor{blue}{0} & \textcolor{blue}{0}  &  \textcolor{blue}{\frac{\text{E}_{\text{P2*F},t}}{\text{E}_{\text{P2,T},t}}} & \textcolor{blue}{\frac{\text{E}_{\text{P2*M},t}}{\text{E}_{\text{P2,T},t}}} & \textcolor{blue}{0} & \textcolor{blue}{0}  & \cdots & \textcolor{blue}{0} & \textcolor{blue}{0} \\
\vdots & \vdots & \vdots & \vdots & \vdots & \vdots & \cdots & \vdots & \vdots \\
\textcolor{blue}{0} & \textcolor{blue}{0} & \textcolor{blue}{0} & \textcolor{blue}{0} & \textcolor{blue}{0} & \textcolor{blue}{0} & \cdots & \textcolor{blue}{\frac{\text{E}_{\text{P47*F},t}}{\text{E}_{\text{P47,T},t}}} & \textcolor{blue}{\frac{\text{E}_{\text{P47*M},t}}{\text{E}_{\text{P47,T},t}}} \\
\textcolor{purple}{1} & \textcolor{purple}{0} & \textcolor{purple}{0} & \textcolor{purple}{0} & \textcolor{purple}{0} & \textcolor{purple}{0} & \cdots & \textcolor{purple}{0} & \textcolor{purple}{0} \\
\textcolor{purple}{0} & \textcolor{purple}{1} & \textcolor{purple}{0} & \textcolor{purple}{0} & \textcolor{purple}{0} & \textcolor{purple}{0} & \cdots & \textcolor{purple}{0} & \textcolor{purple}{0} \\
\textcolor{purple}{0} & \textcolor{purple}{0} & \textcolor{purple}{1} & \textcolor{purple}{0} & \textcolor{purple}{0} & \textcolor{purple}{0} & \cdots & \textcolor{purple}{0} & \textcolor{purple}{0} \\
\textcolor{purple}{0} & \textcolor{purple}{0} & \textcolor{purple}{0} & \textcolor{purple}{1} & \textcolor{purple}{0} & \textcolor{purple}{0} & \cdots & \textcolor{purple}{0} & \textcolor{purple}{0} \\
\vdots & \vdots & \vdots & \vdots & \vdots & \vdots & \cdots & \vdots & \vdots  \\
\textcolor{purple}{0} & \textcolor{purple}{0} & \textcolor{purple}{0} & \textcolor{purple}{0} & \textcolor{purple}{0} & \textcolor{purple}{0} & \cdots  & \textcolor{purple}{1} & \textcolor{purple}{0}\\
\textcolor{purple}{0} & \textcolor{purple}{0} & \textcolor{purple}{0} & \textcolor{purple}{0} & \textcolor{purple}{0} & \textcolor{purple}{0} & \cdots & \textcolor{purple}{0} & \textcolor{purple}{1} \\
\end{array}
\end{footnotesize} \right]}_{\bm{S}_t}
\underbrace{\left[
\begin{footnotesize}
\begin{array}{l}
\text{R}_{\text{P1*F},t} \\
\text{R}_{\text{P1*M},t} \\
\text{R}_{\text{P2*F},t} \\
\text{R}_{\text{P2*M},t} \\
\vdots \\
\text{R}_{\text{P47*F},t} \\
\text{R}_{\text{P47*M},t} \\
\end{array}
\end{footnotesize}
\right]}_{\bm{b}_t}
\]
or $\bm{R}_t = \bm{S}_t\bm{b}_t$, where $\bm{R}_t$ is a vector containing all series at all levels of disaggregation, $\bm{b}_t$ is a vector of the most disaggregated series, and $\bm{S}_t$ shows how the two are connected. 

\cite{HAA+11} considered four hierarchical forecasting methods for univariate time series, namely the top-down, bottom-up, middle-out and optimal-combination methods. Among these four methods, the top-down and middle-out methods rely on a unique hierarchy for assigning disaggregation weights from a higher level series to a lower level series. In contrast, the bottom-up and optimal-combination methods are suitable for forecasting a non-unique group structure. These two methods are reviewed in Sections~\ref{sec:bu} and~\ref{sec:ols}, respectively. Their point-forecast and interval-forecast accuracy comparisons with the independent forecasting method are presented in Sections~\ref{sec:point_compar} and~\ref{sec:interval_compar}, respectively.

\subsection{Bottom-up method}\label{sec:bu}

As the simplest grouped forecasting method, the bottom-up method first generates independent forecasts for each series at the most disaggregated level, and then aggregates these to produce all required forecasts. For example, reverting to the Japanese data, we first generate $h$-step-ahead independent forecasts for the most disaggregated series, namely $\widehat{\bm{b}}_{n+h}=\left[\widehat{\text{R}}_{\text{P}1\ast \text{F},n+h}, \widehat{\text{R}}_{\text{P}1\ast \text{M},n+h},\dots,\widehat{\text{R}}_{\text{P}47\ast \text{F},n+h}, \widehat{\text{R}}_{\text{P}47\ast \text{M},n+h}\right]^{\top}$.

The observed ratios that form the $\bm{S}_t$ summing matrix are forecast using the automatic ARIMA algorithm of \cite{HK08}, when age $x=60$. For age above 60, we assume the exposure-to-risk of age $x+1$ in year $t+1$ will be the same as the exposure-to-risk of age $x$ in year $t$ \citep[see also][]{SH16}. For example, let $p_t = \text{E}_{\text{P}1\ast \text{F},t}/\text{E}_{\text{Japan}\ast \text{T},t}$ be a non-zero element of $\bm{S}_t$. Given that we have observed $\{p_1,\dots,p_n\}$, an $h$-step-ahead forecast $\widehat{p}_{n+h}$ can be obtained. The forecasts of exposure-to-risk are then used to form the matrix $\bm{S}_{n+h}$. Thus we obtain forecasts for all series as
\begin{equation*}
\overline{\bm{R}}_{n+h} = \bm{S}_{n+h}\widehat{\bm{b}}_{n+h},
\end{equation*}
where $\overline{\bm{R}}_{n+h}$ denotes the reconciled forecasts.

The potential improvement in forecast accuracy of the reconciliation methods partially relies on the accurate forecast of the $S$ matrix. Recall that the $S$ matrix includes ratios of forecast exposure-at-risk. Our cohort assumption is reasonable because it allows us to forecast ratios and populate the $S$ matrix. In Sections~\ref{sec:S_point} and~\ref{sec:S_interval}, we compare point-forecast and interval-forecast accuracies between the reconciliation methods, with the forecast $S$ matrix and actual holdout $S$ matrix, respectively.

The bottom-up method performs well when the bottom-level series have a strong signal-to-noise ratio. In contrast, the bottom-up method may lead to inaccurate forecasts of the top-level series, in particular when there are missing or noisy data at the bottom level. 

\subsection{Optimal-combination method}\label{sec:ols}

Instead of considering only the bottom-level series, \cite{HAA+11} proposed the optimal-combination method in which independent forecasts for all series are computed independently, and then the resultant forecasts are reconciled so that they satisfy the aggregation constraints via the summing matrix. The optimal-combination method combines the independent forecasts through linear regression by generating a set of revised forecasts that are as close as possible to the independent forecasts but that also aggregate consistently within the group. The method is derived by expressing the independent forecasts as the response variable of the linear regression.
\begin{equation*}
\widehat{\bm{R}}_{n+h} = \bm{S}_{n+h}\bm{\beta}_{n+h} + \bm{\varepsilon}_{n+h},
\end{equation*}
where $\widehat{\bm{R}}_{n+h}$ is a matrix of $h$-step-ahead independent forecasts for all series, stacked in the same order as for the original data; $\bm{\beta}_{n+h} = \text{E}[\bm{b}_{n+h}|\bm{R}_1,\dots,\bm{R}_n]$ is the unknown mean of the independent forecasts of the most disaggregated series; and $\bm{\varepsilon}_{n+h}$ represents the reconciliation errors. 

To estimate the regression coefficient, \cite{HAA+11} and \cite{Hyndman2016-wp} proposed a weighted least-squares solution, 
\begin{equation*}
\widehat{\bm{\beta}}_{n+h} = \left(\bm{S}_{n+h}^{\top}\bm{W}_h^{-1}\bm{S}_{n+h}\right)^{-1}\bm{S}_{n+h}^{\top}\bm{W}_h^{-1}\widehat{\bm{R}}_{n+h},
\end{equation*}
where $\bm{W}_h$ is a diagonal matrix. Assuming that $\bm{W}_h = k_h\bm{I}$ and $\bm{I}$ denotes identical matrix, then the revised forecasts are given by
\begin{equation*}
\overline{\bm{R}}_{n+h} = \bm{S}_{n+h}\widehat{\bm{\beta}}_{n+h} = \bm{S}_{n+h}\left(\bm{S}_{n+h}^{\top}\bm{S}_{n+h}\right)^{-1}\bm{S}_{n+h}^{\top}\widehat{\bm{R}}_{n+h},
\end{equation*}
where $k_h$ is a constant. These reconciled forecasts are aggregate consistent and involve a combination of all the independent forecasts. They are unbiased because $\text{E}(\widehat{\bm{\beta}}_{n+h})\rightarrow \bm{\beta}_{n+h}$ and $\text{E}(\overline{\bm{R}}_{n+h}) = \bm{S}_{n+h}\bm{\beta}_{n+h}$. 


\subsection{Constructing pointwise and simultaneous prediction intervals}\label{sec:4.4}

As a means of measuring uncertainty associated with point forecasts, prediction intervals based on statistical theory and data on error distributions provide an explicit estimate of the probability that the future realizations lie within a given range. The main sources of uncertainty stem from (1) the error in forecasting principal component scores; (2) the model residuals. As emphasized by \cite{Chatfield93}, it is important to provide interval forecasts as well as point forecasts to be able to
\begin{inparaenum}
\item[(1)] assess future uncertainty level;
\item[(2)] enable different strategies to be planned for the range of possible outcomes;
\item[(3)] compare forecasts from different methods;
\item[(4)] explore different scenarios based on different underlying assumptions.
\end{inparaenum}

To construct pointwise and simultaneous prediction intervals, we adapt the method of \cite{ANH15}. The method can be summarized in the following steps:
\begin{asparaenum}
\item[1)] Using all observations, we compute the $J$-variate score vectors $(\bm{k}_1,\dots,\bm{k}_J)$ and the sample functional principal components $\left[\widehat{b}_1(x),\dots,\widehat{b}_J(x)\right]$. Then, we calculate in-sample point forecasts
\begin{equation}
m_{\xi+h}(x) = \widehat{k}_{\xi+h,1}\widehat{b}_1(x) + \cdots + \widehat{k}_{\xi+h,J}\widehat{b}_J(x),
\end{equation}
where $\left(\widehat{k}_{\xi+h,1},\dots,\widehat{k}_{\xi+h,J}\right)$ are the elements of the $h$-step-ahead prediction obtained from $\left(\widehat{\bm{k}}_1,\dots,\widehat{\bm{k}}_J\right)$ by a means of a univariate or multivariate time-series forecasting method, for $\xi\in\{J,\dots,n-h\}$.
\item[2)] With the in-sample point forecasts, we calculate the in-sample point-forecast errors
\begin{equation}
\widehat{\epsilon}_{\omega}(x) = m_{\xi+h}(x) - \widehat{m}_{\xi+h}(x),
\end{equation}
where $\omega \in\{1,2,\dots,M\}$ and $M=n-h-J+1$.
\item[3)] Based on these in-sample forecast errors, we can sample with replacement to obtain a series of bootstrapped forecast errors, from which we obtain lower and upper prediction intervals, denoted by $\gamma^{l}(x)$ and $\gamma^{u}(x)$, respectively. We then seek a tuning parameter $\psi_{\alpha}$ such that $\alpha\times 100\%$ of the residual functions satisfy
\begin{equation}
\psi_{\alpha}\times \gamma^l(x)\leq \widehat{\epsilon}_{\omega}(x)\leq \psi_{\alpha}\times \gamma^u(x).
\end{equation}
The residuals $\widehat{\epsilon}_1(x),\dots,\widehat{\epsilon}_M(x)$ are then expected to be approximately stationary and by the law of large numbers, to satisfy
\begin{align*}
\frac{1}{M}\sum_{\omega=1}^M\mathds{1}\left(\psi_{\alpha}\times \gamma^l(x)\leq \widehat{\epsilon}_{\omega}(x)\leq \psi_{\alpha}\times \gamma^u(x)\right) \approx \text{Pr}\left[\psi_{\alpha}\times \gamma^l(x)\leq m_{n+h}(x) - \widehat{m}_{n+h}(x) \leq \psi_{\alpha}\times \gamma^u(x)\right].
\end{align*}
\end{asparaenum}
Note that \cite{ANH15} calculate the standard deviation function of $\left[\widehat{\epsilon}_1(x),\dots,\widehat{\epsilon}_M(x)\right]$, which leads to a parametric approach of constructing prediction intervals. Instead, we consider the nonparametric approach of \cite{Shang16c}, which allows us to reconcile bootstrapped forecasts among different functional time series in a hierarchy. Step 3) can easily be extended to pointwise prediction intervals, where we determine $\psi_{\alpha}$ such that $\alpha\times 100\%$ of the residual data points satisfy
\begin{equation}
\psi_{\alpha}\times \gamma^l(x_i)\leq \widehat{\epsilon}_{\omega}(x_i) \leq \psi_{\alpha}\times \gamma^u(x_i).
\end{equation}
Then, the $h$-step-ahead pointwise prediction intervals are given as
\begin{equation}
\psi_{\alpha}\times \gamma^l(x_i) \leq m_{n+h}(x_i) - \widehat{m}_{n+h}(x_i) \leq \psi_{\alpha}\times\gamma^u(x_i), 
\end{equation}
where $i$ symbolizes the discretized data points. A simultaneous confidence interval will generally be wider than a pointwise confidence interval with the same coverage probability.

\section{Results -- point forecasts}

\subsection{Functional time series model fitting}\label{sec:fitting}

For the national and subnational mortality rates, we examine the goodness-of-fit of the functional time series model to the smoothed data. The number of retained components in the functional principal component decomposition is determined by explaining at least 95\% of the total variation. We present and interpret the first component for the female mortality series in Hokkaido as an illustration.

\begin{figure}[!htbp]
\centering
\includegraphics[width=17cm]{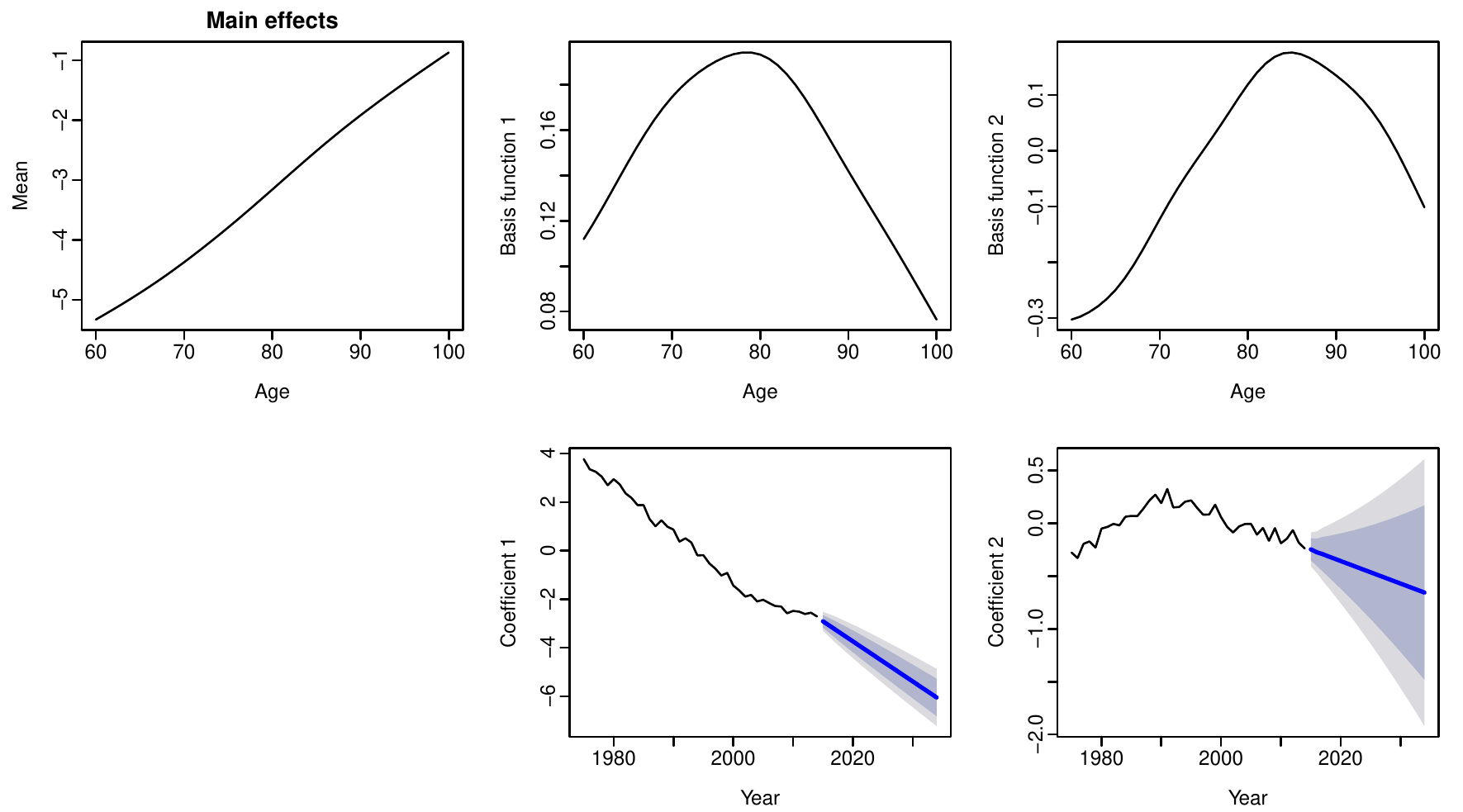}
\caption{Functional principal component decomposition for the female mortality data in Hokkaido. In the bottom panel, the solid blue line represents the point forecasts of scores, where the dark and light grey regions represent the 80\% and 95\% pointwise prediction intervals.}\label{fig:fts_model}
\end{figure}

In the first column of Figure~\ref{fig:fts_model}, we present the average of female log mortality rates. In the first row of Figure~\ref{fig:fts_model}, we also present the first functional principal component, which accounts for 98.9\% of the total variation. The functional principal component models different movements in mortality rates. By inspecting the peaks, it models the mortality at approximately age 80. Given that the principal component scores are surrogates of the original functional time series, they are forecast to continue to decrease over the next 20 years.

\begin{figure}[!htbp]
\centering
\subfloat[Fitted values]
{\includegraphics[width=8.2cm]{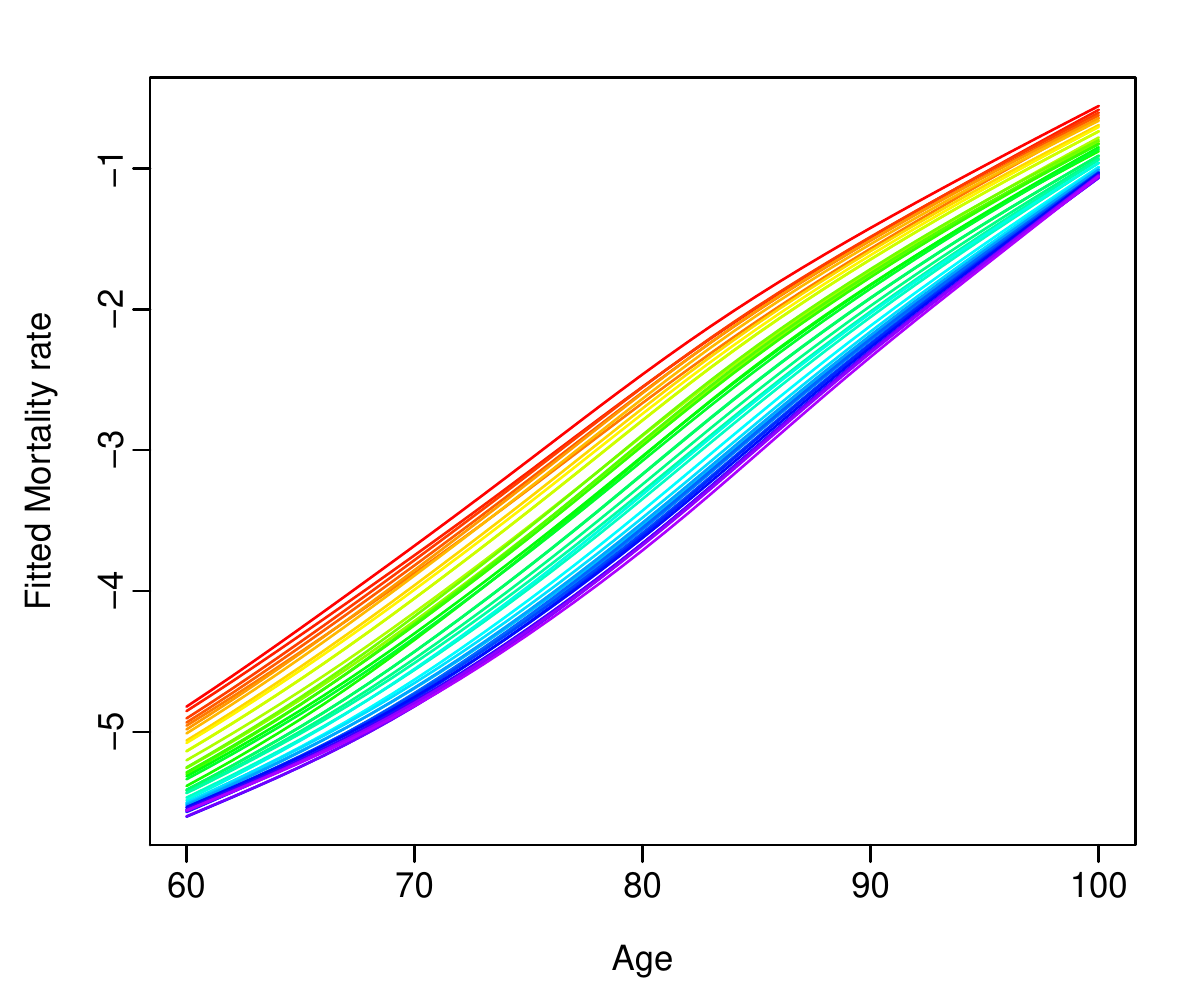}\label{fig:resi_a}}
\quad
\subfloat[Residual values]
{\includegraphics[width=8.2cm]{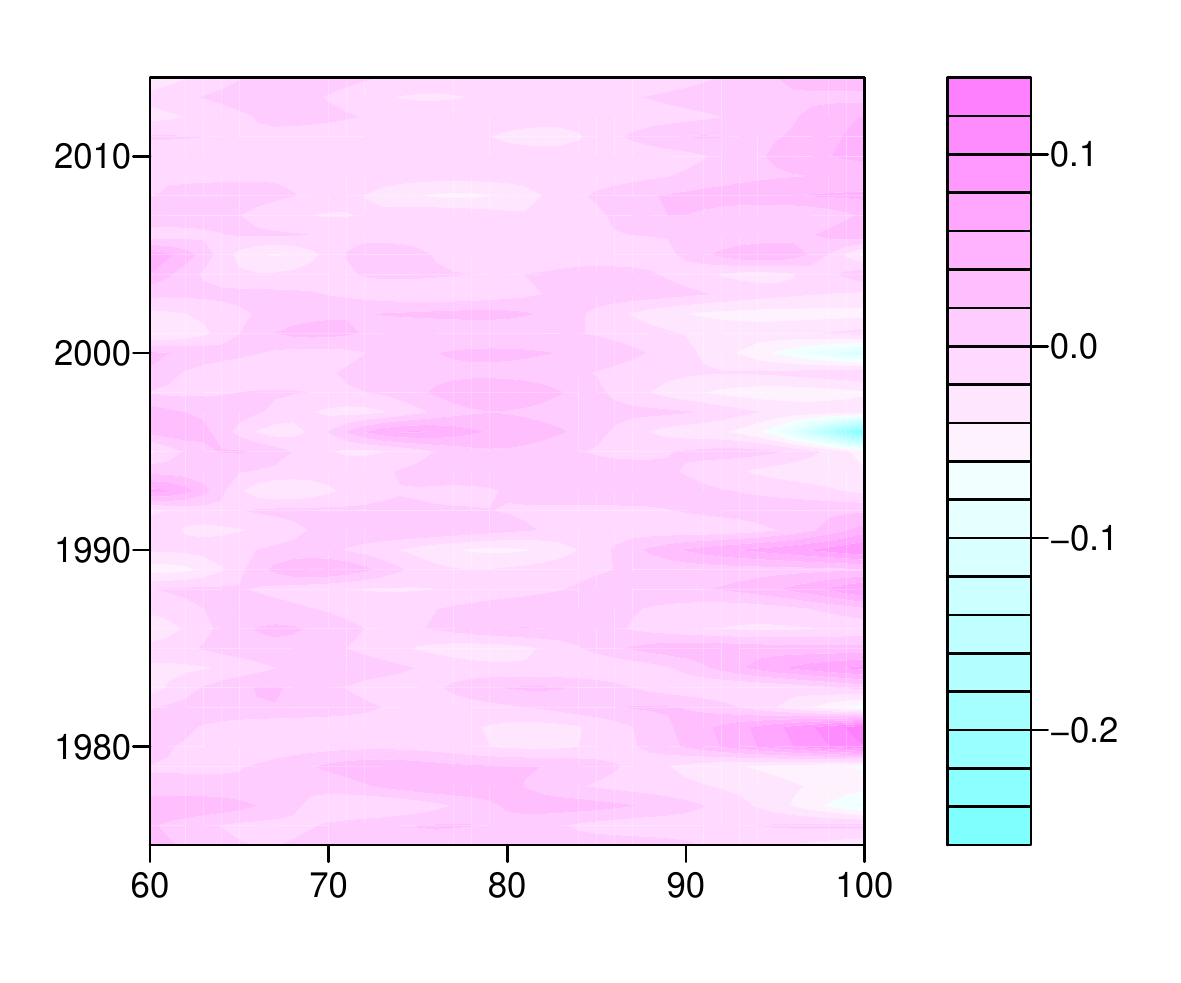}\label{fig:resi_b}}
\caption{Functional time series model fitting and residual filled contour plots.}\label{fig:resi}
\end{figure}

In Figure~\ref{fig:resi_a}, we present the functional time series model fit to the smoothed data. The difference between the fitted and smoothed data (i.e., residuals) is highlighted in a filled contour plot in Figure~\ref{fig:resi_b}. 

In addition to the graphical display, we measure goodness-of-fit via a functional version of the $R^2$ criterion. It is given as
\begin{equation}
R^2 = 1 - \frac{\int_{x\in \mathcal{I}} \sum_{t=1}^n\left[\exp^{m_t(x)} - \exp^{\widehat{f}_t(x)}\right]^2dx}{\int_{x\in \mathcal{I}} \sum^n_{t=1}\left[\exp^{m_t(x)} - \exp^{\overline{m}(x)}\right]^2dx}, \label{eq:R2}
\end{equation}
where $m_t(x)$ denotes the observed age-specific log mortality rates, $\widehat{f}_t(x)$ denotes the fitted age-specific log mortality rates. The larger the $R^2$ value is, the better is the goodness-of-fit by the functional time series model. It is possible for the $R^2$ criterion to take negative values. A negative $R^2$ value implies that the fitted model may not well explain the raw data that are likely to contain a large amount of measurement errors. From a negative $R^2$ value, we can quantify the amount of measurement errors exhibited in a data set and the degree of smoothing that the raw mortality data require.

\begin{figure}[!htbp]
\centering
\includegraphics[width=8.5cm]{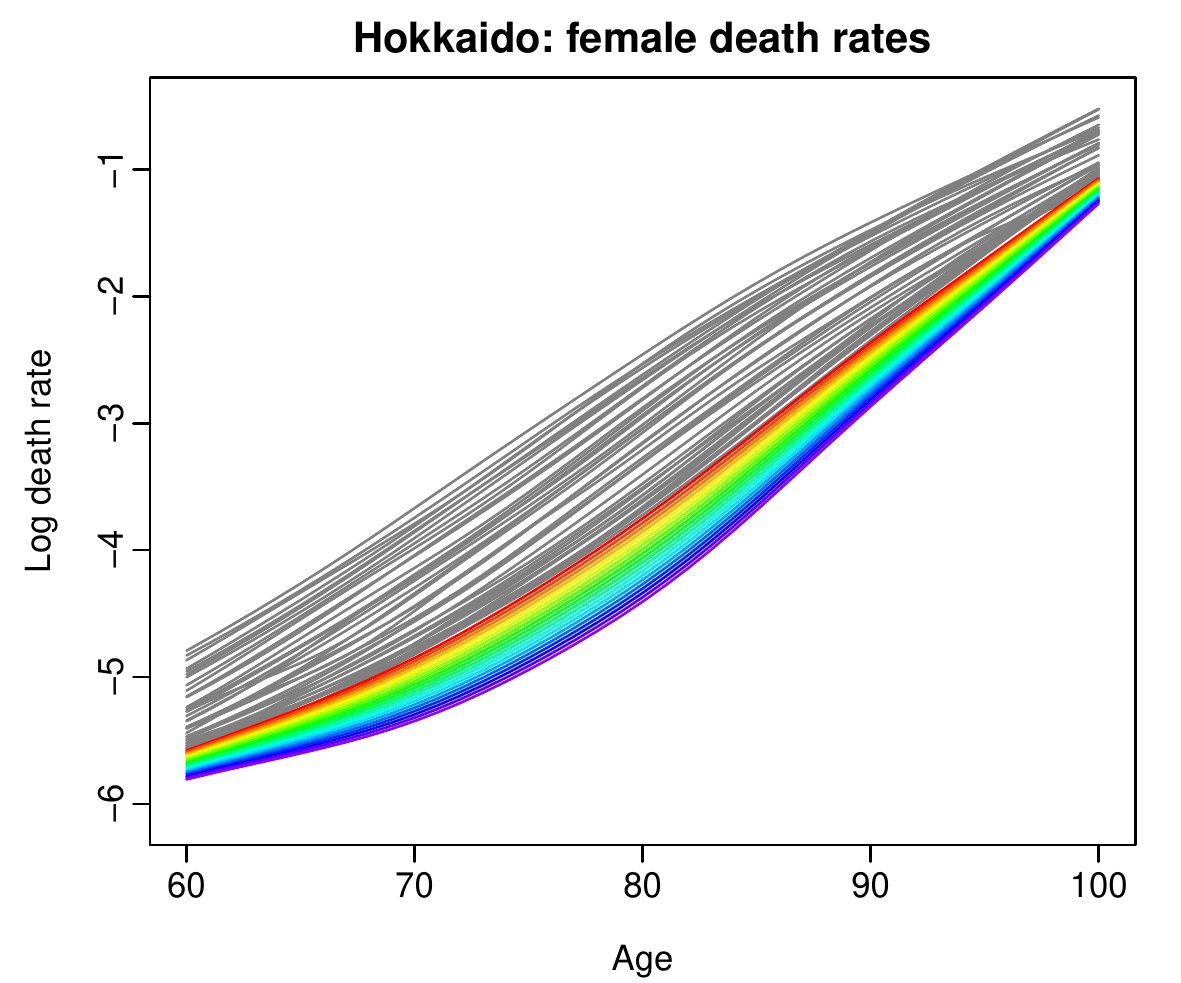}
\caption{Point forecasts of age-specific mortality rates from 2015 to 2034. The historical functional time series is shown in gray, and the forecasts are highlighted in rainbow color.}\label{fig:smooth_point}
\end{figure}

Based on the historical mortality from 1975 to 2014, we produce the point forecasts of age-specific mortality rates from 2015 to 2034. As shown in Figure~\ref{fig:smooth_point}, the mortality rates are continuing to decline, particularly for the population over 60.

Due to space limitations, we cannot present the functional time series model fitting and forecasts for each subnational group. However, in Table~\ref{tab:component_number}, we report the retained number of components determined by explaining at least 95\% of total variation, and the goodness-of-fit of the functional time series model as measured by the $R^2$ criterion in~\eqref{eq:R2}. This retained number of components is used in the functional time series model for fitting each series. 

\begin{table}[!htbp] \centering 
\tabcolsep 0.09in
\caption{Number of retained functional principal components $J$ and the goodness-of-fit as measured by the $R^2$ for each national and subnational female, male and total mortality rates in Japan.}\label{tab:component_number}
\begin{tabular}{@{} llrlrlrllrlrlr} 
\toprule
& \multicolumn{2}{c}{Female} & \multicolumn{2}{c}{Male} & \multicolumn{2}{c}{Total} & & \multicolumn{2}{c}{Female} & \multicolumn{2}{c}{Male} & \multicolumn{2}{c}{Total} \\
Prefecture  & $J$ & $R^2$ & $J$ & $R^2$ & $J$ & $R^2$ & Prefecture  & $J$ & $R^2$ & $J$ & $R^2$ & $J$ & $R^2$ \\
\midrule
Japan & $1$ & $0.95$ & $1$ & $0.82$ & $1$ & $0.96$ & Mie & $1$ & $0.30$ & $2$ & $0.11$ & $1$ & $0.45$ \\ 
Hokkaido & $1$ & $0.47$ & $1$ & $0.26$ & $1$ & $0.64$ & Shiga & $1$ & $0.11$ & $1$ & $-0.11$ & $1$ & $0.31$ \\  
Aomori & $1$ & $0.12$ & $2$ & $0.00$ & $1$ & $0.24$ & Kyoto & $1$ & $0.47$ & $1$ & $0.05$ & $1$ & $0.52$ \\  
Iwate & $1$ & $0.21$ & $2$ & $0.12$ & $1$ & $0.36$ & Osaka & $1$ & $0.66$ & $1$ & $0.23$ & $1$ & $0.72$ \\  
Miyagi & $1$ & $0.15$ & $2$ & $0.01$ & $1$ & $0.28$ & Hyogo & $1$ & $0.60$ & $1$ & $0.23$ & $1$ & $0.68$ \\  
Akita & $1$ & $0.22$ & $2$ & $-0.17$ & $1$ & $0.19$ & Nara & $1$ & $0.25$ & $1$ & $-0.11$ & $1$ & $0.30$ \\  
Yamagata & $1$ & $0.16$ & $2$ & $-0.02$ & $1$ & $0.25$ & Wakayama & $1$ & $0.37$ & $2$ & $-0.15$ & $1$ & $0.45$ \\  
Fukushima & $1$ & $0.37$ & $1$ & $0.02$ & $1$ & $0.41$ & Tottori & $1$ & $0.10$ & $2$ & $-0.12$ & $1$ & $0.24$ \\  
Ibaraki & $1$ & $0.50$ & $2$ & $0.16$ & $1$ & $0.55$ & Shimane & $1$ & $0.36$ & $1$ & $-0.08$ & $1$ & $0.43$ \\  
Tochigi & $1$ & $0.33$ & $2$ & $-0.06$ & $1$ & $0.39$ & Okayama & $1$ & $0.42$ & $1$ & $-0.03$ & $1$ & $0.52$ \\  
Gunma & $1$ & $0.27$ & $1$ & $0.00$ & $1$ & $0.41$ & Hiroshima & $1$ & $0.60$ & $2$ & $0.17$ & $1$ & $0.68$ \\  
Saitama & $1$ & $0.44$ & $1$ & $0.06$ & $1$ & $0.48$ & Yamaguchi & $1$ & $0.35$ & $2$ & $0.12$ & $1$ & $0.44$ \\ 
Chiba & $1$ & $0.54$ & $2$ & $-0.08$ & $1$ & $0.61$ & Tokushima & $1$ & $0.35$ & $2$ & $0.12$ & $1$ & $0.44$ \\  
Tokyo & $1$ & $0.58$ & $1$ & $0.32$ & $1$ & $0.66$ & Kagawa & $1$ & $0.22$ & $2$ & $0.03$ & $1$ & $0.29$ \\  
Kanagawa & $1$ & $0.54$ & $1$ & $0.01$ & $1$ & $0.54$ & Ehime & $1$ & $0.26$ & $2$ & $0.19$ & $1$ & $0.48$ \\  
Niigata & $1$ & $0.49$ & $1$ & $0.15$ & $1$ & $0.57$ & Kochi & $1$ & $0.40$ & $2$ & $0.01$ & $1$ & $0.43$ \\  
Toyama & $1$ & $0.18$ & $2$ & $0.07$ & $1$ & $0.32$ & Fukuoka & $1$ & $0.55$ & $1$ & $0.18$ & $1$ & $0.62$ \\  
Ishikawa & $1$ & $0.02$ & $2$ & $-0.14$ & $1$ & $0.19$ & Saga & $1$ & $0.20$ & $1$ & $-0.12$ & $1$ & $0.30$ \\  
Fukui & $1$ & $0.23$ & $1$ & $-0.12$ & $1$ & $0.33$ & Nagasaki & $1$ & $0.31$ & $2$ & $-0.03$ & $1$ & $0.40$ \\  
Yamanashi & $1$ & $0.28$ & $2$ & $-0.11$ & $1$ & $0.31$ & Kumamoto & $1$ & $0.48$ & $1$ & $0.10$ & $1$ & $0.57$ \\  
Nagano & $1$ & $0.34$ & $1$ & $0.08$ & $1$ & $0.38$ & Oita & $1$ & $0.32$ & $1$ & $0.18$ & $1$ & $0.49$ \\  
Gifu & $1$ & $0.42$ & $2$ & $0.15$ & $1$ & $0.51$ & Miyazaki & $1$ & $0.27$ & $1$ & $0.09$ & $1$ & $0.39$ \\  
Shizuoka & $1$ & $0.47$ & $2$ & $0.21$ & $1$ & $0.59$ & Kagoshima & $1$ & $0.46$ & $1$ & $0.10$ & $1$ & $0.55$ \\  
Aichi & $1$ & $0.49$ & $1$ & $0.15$ & $1$ & $0.58$ & Okinawa & $1$ & $0.47$ & $2$ & $0.08$ & $1$ & $0.51$ \\ 
\bottomrule
\end{tabular} 
\end{table}

\subsection{Point-forecast evaluation}\label{sec:point_eval}

An expanding window analysis of a time-series model is commonly used to assess model and parameter stability over time, and prediction accuracy. The expanding window analysis assesses the constancy of a model's parameter by computing parameter estimates and their resultant forecasts over an expanding window of a fixed size through the sample \citep[For details,][pp. 313-314]{ZW06}. Using the first 25 observations from 1975 to 1999 in the Japanese age-specific mortality rates, we produce one- to 15-step-ahead point forecasts. Through a rolling-window approach, we re-estimate the parameters in the time-series forecasting models using the first 26 observations from 1975 to 2000. Forecasts from the estimated models are then produced for one- to 14-step-ahead. We iterate this process by increasing the sample size by one year until reaching the end of the data period in 2014. This process produces 15 one-step-ahead forecasts, 14 two-step-ahead forecasts, $\dots$, and one 15-step-ahead forecast. We compare these forecasts with the holdout samples to determine the out-of-sample point-forecast accuracy.

To evaluate the point-forecast accuracy, we consider the MAFE and RMSFE. These criteria measure how close the forecasts are in comparison to the actual values of the variable being forecast, regardless of the direction of forecast errors. For each series $k$, these error measures can be written as
\begin{align*}
\text{MAFE}_k(h) &=\frac{1}{41\times (16-h)}\sum^{15}_{\varsigma=h}\sum^{41}_{j=1}\left|m_{n+\varsigma}^k(x_j) - \widehat{m}_{n+\varsigma}^k(x_j)\right|, \\
\text{RMSFE}_k(h) &= \sqrt{\frac{1}{41\times (16-h)}\sum^{15}_{\varsigma=h}\sum^{41}_{j=1}\left[m_{n+\varsigma}^k(x_j) - \widehat{m}_{n+\varsigma}^k(x_j)\right]^2},
\end{align*}
where $m_{n+\varsigma}^k(x_j)$ denotes the actual holdout sample for the $j^{\text{th}}$ age and $\varsigma^{\text{th}}$ curve in the $k^{\text{th}}$ series, while $\widehat{m}_{n+\varsigma}^k(x_j)$ denotes the point forecasts for the holdout sample.

By averaging MAFE$_k(h)$ and RMSFE$_k(h)$ across the number of series within each level of disaggregation, we obtain an overall assessment of the point-forecast accuracy for each level within the collection of series, denoted by MAFE$(h)$ and RMSFE$(h)$. These error measures are defined as
\begin{align*}
\text{MAFE}(h) &= \frac{1}{m_k}\sum^{m_k}_{k=1}\text{MAFE}_k(h), \\
\text{RMSFE}(h) &= \frac{1}{m_k}\sum^{m_k}_{k=1}\text{RMSFE}_k(h),
\end{align*}
where $m_k$ denotes the number of series at the $k^{\text{th}}$ level of disaggregation, for $k=1,\dots,K$.

For 15 different forecast horizons, we consider two summary statistics to evaluate overall point-forecast accuracy among the methods for national and subnational mortality forecasts. The summary statistics chosen are the mean and median values due to their suitability for handling squared and absolute errors \citep{Gneiting11}. These error measures are given by
\begin{align*}
\text{Mean (RMSFE)} &= \frac{1}{15}\sum^{15}_{h=1}\text{RMSFE}(h), \\
\text{Median (MAFE)} &= \text{MAFE}[8],
\end{align*}
where $[8]$ denotes the $8^{\text{th}}$ term after ranking MAFE$(h)$ for $h=1,\dots,15$ from smallest to largest.

\subsection{Comparison of point-forecast accuracy}\label{sec:point_compar}

\begin{table}[!htbp] 
\centering 
  \caption{MAFEs and RMSFEs $(\times 100)$ in the holdout sample between the functional time series and Lee--Carter methods applied to the Japanese mortality rates. The bold entries highlight the method that gives most accurate summary statistics of the forecasts for each level of the hierarchy. The forecast errors have been multiplied by 100 to keep two decimal places.} \label{tab:HU_LC} 
  \tabcolsep 0.063in
\begin{tabular}{@{\extracolsep{5pt}} lccccc|ccccc@{}} 
\toprule 
& \multicolumn{5}{c|}{MAFE} & \multicolumn{5}{c}{RMSFE} \\ 
Level & $h=1$ & 5 & 10 & 15 & Median & $h=1$ & 5 & 10 & 15 & Mean \\ 
\midrule 
\multicolumn{6}{l|}{\hspace{-.1in} \textit{Functional time series method}} \\
Total & $0.32$ & $0.52$ & $0.74$ & $0.90$ & $\textBF{0.65}$ & $0.58$ & $0.94$ & $1.35$ & $1.47$ & $\textBF{1.13}$ \\  
Sex & $0.36$ & $0.58$ & $0.85$ & $0.96$ & $\textBF{0.73}$ & $0.74$ & $1.14$ & $1.69$ & $1.76$ & $\textBF{1.40}$ \\  
Region & $0.42$ & $0.57$ & $0.80$ & $0.91$ & $\textBF{0.71}$ & $0.90$ & $1.14$ & $1.56$ & $1.61$ & $\textBF{1.34}$ \\ 
Region + Sex & $0.53$ & $0.66$ & $0.92$ & $1.02$ & $\textBF{0.80}$ & $1.21$ & $1.44$ & $1.94$ & $1.99$ & $\textBF{1.68}$ \\  
Prefecture & $0.61$ & $0.70$ & $0.87$ & $0.96$ & $\textBF{0.81}$ & $1.34$ & $1.39$ & $1.66$ & $1.70$ & $\textBF{1.53}$ \\  
Prefecture + Sex & $0.96$ & $0.97$ & $1.12$ & $1.17$ & $\textBF{1.05}$ & $2.24$ & $2.15$ & $2.34$ & $2.32$ & $\textBF{2.26}$ \\  
\\
\multicolumn{6}{l}{\hspace{-.1in} \textit{Lee--Carter method}} \\
Total  & $0.30$ & $0.58$ & $0.97$ & $0.54$ & $0.72$ & $0.59$ & $1.06$ & $1.67$ & $0.99$ & $1.25$ \\   
Sex  & $0.35$ & $0.67$ & $1.06$ & $0.62$ & $0.80$ & $0.76$ & $1.30$ & $1.98$ & $1.28$ & $1.52$ \\    
Region & $0.39$ & $0.61$ & $0.97$ & $0.65$ & $0.73$ & $0.81$ & $1.17$ & $1.74$ & $1.29$ & $1.38$ \\    
Region + Sex & $0.53$ & $0.78$ & $1.15$ & $0.86$ & $0.90$ & $1.21$ & $1.68$ & $2.37$ & $2.00$ & $1.95$ \\  
Prefecture & $0.64$ & $0.81$ & $1.15$ & $0.96$ & $0.96$ & $1.43$ & $1.72$ & $2.30$ & $2.23$ & $2.00$ \\  
Prefecture + Sex & $1.15$ & $1.42$ & $1.88$ & $1.85$ & $1.67$ & $2.91$ & $3.52$ & $4.56$ & $4.67$ & $4.04$ \\ 
\bottomrule 
\end{tabular} 
\end{table} 

Averaging over all the series at each level of a hierarchy, Table~\ref{tab:HU_LC} presents MAFE$(h)$ and RMSFE$(h)$ between the Lee--Carter and functional time series methods. As measured by the MAFE and RMSFE, the functional time series method produces more accurate point forecasts than the ones obtained using the Lee--Carter method, at each level of the hierarchy. The superior forecast accuracy of the functional time series model over the Lee--Carter model stems from two sources: 
\begin{inparaenum}
\item[(1)] a smoothing technique is implemented to remove any noise in the data series, particularly at older ages; 
\item[(2)] more than one component is used to achieve improved model fitting.
\end{inparaenum}

\begin{table}[!htbp] \centering 
  \caption{MAFEs and RMSFEs $(\times 100)$ in the holdout sample among the independent (base) forecasting, bottom-up and optimal-combination methods, using the forecast $S$ matrix.}\label{tab:mae} 
    \tabcolsep 0.032in
\begin{tabular}{@{\extracolsep{5pt}} ll|ccccc|ccccc@{}} 
\toprule 
& & \multicolumn{5}{c|}{MAFE} & \multicolumn{5}{c}{RMSFE} \\ 
Level & Method & $h=1$ & 5 & 10 & 15 & Median & 1 & 5 & 10 & 15 & Mean \\ 
\midrule
Total & Base & $0.32$ & $0.52$ & $0.74$ & $0.90$ & $0.65$ & $0.58$ & $0.94$ & $1.35$ & $1.47$ & $1.13$ \\ 
 & BU & $0.32$ & $0.43$ & $0.52$ & $0.47$ & $\textBF{0.44}$ & $0.63$ & $0.80$ & $1.00$ & $0.93$ & $\textBF{0.86}$ \\ 
 & OLS & $0.28$ & $0.43$ & $0.59$ & $0.68$ & $0.51$ & $0.53$ & $0.82$ & $1.16$ & $1.24$ & $0.97$ \\ 
\midrule
 Sex & Base & $0.36$ & $0.58$ & $0.85$ & $0.96$ & $0.73$ & $0.74$ & $1.14$ & $1.69$ & $1.76$ & $1.40$ \\ 
& BU & $0.36$ & $0.54$ & $0.80$ & $0.93$ & $\textBF{0.67}$ & $0.70$ & $1.00$ & $1.52$ & $1.62$ & $\textBF{1.25}$ \\ 
 & OLS & $0.32$ & $0.57$ & $0.92$ & $1.14$ & $0.76$ & $0.64$ & $1.07$ & $1.71$ & $1.92$ & $1.39$ \\ 
\midrule
 Region & Base & $0.42$ & $0.57$ & $0.80$ & $0.91$ & $0.71$ & $0.90$ & $1.14$ & $1.56$ & $1.61$ & $1.34$ \\ 
 & BU & $0.41$ & $0.48$ & $0.59$ & $0.57$ & $\textBF{0.52}$ & $0.86$ & $0.95$ & $1.16$ & $1.13$ & $\textBF{1.03}$ \\ 
 & OLS & $0.37$ & $0.48$ & $0.64$ & $0.72$ & $0.56$ & $0.76$ & $0.96$ & $1.29$ & $1.35$ & $1.11$ \\ 
\midrule
 Region + Sex & Base & $0.53$ & $0.66$ & $0.92$ & $1.02$ & $0.80$ & $1.21$ & $1.44$ & $1.94$ & $1.99$ & $1.68$ \\ 
 & BU & $0.51$ & $0.63$ & $0.89$ & $1.01$ & $\textBF{0.77}$ & $1.12$ & $1.29$ & $1.75$ & $1.85$ & $\textBF{1.52}$ \\ 
 & OLS & $0.48$ & $0.66$ & $0.98$ & $1.19$ & $0.82$ & $1.06$ & $1.34$ & $1.91$ & $2.11$ & $1.64$ \\ 
\midrule
 Prefecture & Base & $0.61$ & $0.70$ & $0.88$ & $0.96$ & $0.81$ & $1.33$ & $1.39$ & $1.67$ & $1.70$ & $1.53$ \\ 
 & BU & $0.62$ & $0.65$ & $0.72$ & $0.67$ & $\textBF{0.66}$ & $1.40$ & $1.36$ & $1.45$ & $1.35$ & $\textBF{1.39}$ \\ 
 & OLS & $0.60$ & $0.64$ & $0.76$ & $0.78$ & $0.72$ & $1.33$ & $1.35$ & $1.54$ & $1.53$ & $1.45$ \\ 
\midrule
 Prefecture + Sex & Base & $0.96$ & $0.97$ & $1.12$ & $1.17$ & $\textBF{1.05}$ & $2.24$ & $2.15$ & $2.34$ & $2.32$ & $\textBF{2.26}$ \\ 
 & BU & $0.96$ & $0.97$ & $1.12$ & $1.17$ & $\textBF{1.05}$ & $2.24$ & $2.15$ & $2.34$ & $2.32$ & $\textBF{2.26}$ \\ 
 & OLS & $0.94$ & $0.99$ & $1.19$ & $1.30$ & $1.10$ & $2.20$ & $2.18$ & $2.46$ & $2.50$ & $2.34$ \\ 
\bottomrule
\end{tabular} 
\end{table} 

Given that the functional time series method outperforms the Lee--Carter method, we evaluate and compare MAFE$(h)$ and RMSFE$(h)$ in Table~\ref{tab:mae}, among the independent functional time series forecasting, its corresponding bottom-up and optimal-combination methods, for each level within the Japanese data hierarchy. As the forecast horizon increases, the bottom-up method generally gives the most accurate forecasts at the national and subnational levels for the total series, except when total mortality rates are disaggregated by sex. Based on averaged forecast errors, the bottom-up method performs the best at each level of the hierarchy, while it reconciles forecasts with respect to the group structure.

\subsection{Influence of the $S$ matrix on point-forecast accuracy}\label{sec:S_point}

The potential improvement in point-forecast accuracy in the reconciliation methods relies partially on the accurate forecast of the $S$ matrix. Recall that the $S$ matrix includes ratios of forecast exposure-at-risk. To forecast the exposure-at-risk, we use the automatic ARIMA method to model and forecast exposure-at-risk at age 60 on the logarithmic scale. By taking the exponential back-transformation, forecast exposure-at-risk on the original scale is obtained. For age above 60, we assume the exposure-to-risk of age $x+1$ in year $t+1$ will be the same as the exposure-to-risk of age $x$ in year $t$. In Table~\ref{tab:actual_S_point}, we present the MAFEs and RMSFEs between the reconciliation methods with the holdout $S$ matrix.

\begin{table}[!htbp]
\centering
\caption{MAFEs and RMSFEs ($\times 100$) in the holdout sample between the bottom-up and optimal-combination methods, using the actual holdout $S$ matrix.}\label{tab:actual_S_point}
    \tabcolsep 0.063in
\begin{tabular}{@{}ll|ccccc|ccccc@{}}
  \toprule
  & & \multicolumn{5}{c|}{MAFE} & \multicolumn{5}{c}{RMSFE} \\
Level & Method & $h=1$ & 5 & 10 & 15 & Median & $h=1$ & 5 & 10 & 15 & Mean \\ 
  \midrule
 Total & BU & 0.32 & 0.51 & 0.75 & 0.94 & 0.63 & 0.63 & 0.96 & 1.44 & 1.65 & 1.20 \\ 
 & OLS & 0.28 & 0.50 & 0.78 & 0.99 & 0.65 & 0.53 & 0.96 & 1.48 & 1.71 & 1.22 \\  
\cmidrule{2-12}
   Sex & BU &  0.36 & 0.55 & 0.84 & 0.98 & 0.70 & 0.70 & 1.05 & 1.61 & 1.74 & 1.32 \\ 
   & OLS & 0.32 & 0.56 & 0.87 & 1.04 & 0.72 & 0.64 & 1.08 & 1.67 & 1.84 & 1.36 \\ 
\cmidrule{2-12}
   Region & BU & 0.41 & 0.55 & 0.79 & 0.95 & 0.68 & 0.86 & 1.07 & 1.51 & 1.67 & 1.30 \\ 
   & OLS & 0.37 & 0.53 & 0.80 & 0.99 & 0.66 & 0.76 & 1.05 & 1.53 & 1.72 & 1.30 \\ 
\cmidrule{2-12}
   Region + Sex & BU &  0.51 & 0.64 & 0.90 & 1.02 & 0.77 & 1.12 & 1.29 & 1.77 & 1.87 & 1.54 \\ 
   & OLS & 0.48 & 0.63 & 0.90 & 1.07 & 0.77 & 1.06 & 1.30 & 1.80 & 1.94 & 1.56 \\ 
\cmidrule{2-12}
   Prefecture & BU & 0.62 & 0.69 & 0.87 & 0.97 & 0.79 & 1.40 & 1.43 & 1.70 & 1.79 & 1.58 \\ 
   & OLS & 0.60 & 0.68 & 0.88 & 1.00 & 0.80 & 1.33 & 1.40 & 1.71 & 1.82 & 1.58 \\ 
\cmidrule{2-12}
   Prefecture + Sex & BU & 0.96 & 0.97 & 1.12 & 1.17 & 1.05 & 2.24 & 2.15 & 2.34 & 2.32 & 2.26 \\ 
   & OLS & 0.94 & 0.97 & 1.13 & 1.20 & 1.06 & 2.20 & 2.15 & 2.37 & 2.36 & 2.27 \\   
   \bottomrule
\end{tabular}
\end{table}

\vspace{.3in}

For the reconciliation methods, more accurate forecasts can generally be obtained using the forecast $S$ matrix. This is because the reconciliation methods rely on the forecast mortality counts. For example, we forecast the age-specific mortality rate in prefecture Okinawa, the forecast is then multiplied by a ratio involving the exposure-at-risk between Okinawa and Japan. The mortality forecasts in Okinawa contribute partially to the Japanese national mortality rate forecasts. Even if we use the actual ratios of exposure-at-risk, the forecast errors may stem from the forecast mortality rates.

\section{Results -- interval forecasts}\label{sec:interval_section}

\subsection{Interval-forecast evaluation}\label{sec:interval_eval}

To evaluate pointwise interval-forecast accuracy, we utilize the interval score of \cite{GR07}. For each year in the forecasting period, the $h$-step-ahead prediction intervals are calculated at the $100(1-\alpha)\%$ nominal coverage probability. We consider the common case of the symmetric $100(1-\alpha)\%$ prediction intervals, with lower and upper bounds that are predictive quantiles at $\alpha/2$ and $1-\alpha/2$, denoted by $\widehat{m}_{\zeta+h|\zeta}^l(x_i)$ and $\widehat{m}_{\zeta+h|\zeta}^u(x_i)$. As defined by \cite{GR07}, a scoring rule for the interval forecasts at time point $m_{\zeta+h}(x_j)$ is
\begin{align*}
S_{\alpha}\left[\widehat{m}^l_{\zeta+h|\zeta}(x_j), \widehat{m}^u_{\zeta+h|\zeta}(x_j); m_{\zeta+h}(x_j)\right] = \left[\widehat{m}^u_{\zeta+h|\zeta}(x_j) - \widehat{m}^l_{\zeta+h|\zeta}(x_j)\right] + \frac{2}{\alpha}\left[\widehat{m}^l_{\zeta+h|\zeta}(x_j) - m_{\zeta+h}(x_j)\right] \\
\mathds{1}\left\{m_{\zeta+h}(x_j) < \widehat{m}^l_{\zeta+h|\zeta}(x_j)\right\} + \frac{2}{\alpha}\left[m_{\zeta+h}(x_j) - \widehat{m}^u_{\zeta+h|\zeta}(x_j)\right] \mathds{1}\left\{m_{\zeta+h}(x_j) > \widehat{m}^u_{\zeta+h|\zeta}(x_j)\right\},
\end{align*}
where $\mathds{1}\{\cdot\}$ represents the binary indicator function, and $\alpha$ denotes the level of significance, customarily $\alpha = 0.2$. The interval score rewards a narrow prediction interval, if and only if the true observation lies within the prediction interval. The optimal interval score is achieved when $m_{n+h}(x_j)$ lies between $\widehat{m}_{n+h|n}^l(x_j)$ and $\widehat{m}_{n+h|n}^u(x_j)$, and the distance between $\widehat{m}_{n+h|n}^l(x_j)$ and $\widehat{m}_{n+h|n}^u(x_j)$ is minimal. 

For different ages and years in the forecasting period, the mean interval score is defined by
\begin{equation}
\overline{S}_{\alpha}(h) = \frac{1}{41\times (16-h)}\sum^{15}_{\varsigma=h}\sum^{41}_{j=1}S_{\alpha,\varsigma}\left[\widehat{m}_{n+h|n}^l(x_j), \widehat{m}_{n+h|n}^u(x_j); m_{n+h}(x_j)\right],
\end{equation}
where $S_{\alpha,\varsigma}\left[\widehat{m}_{n+h|n}^l(x_j), \widehat{m}_{n+h|n}^u(x_j); m_{n+h}(x_j)\right]$ denotes the interval score at the $\varsigma^{\text{th}}$ curve in the forecasting period.

For 15 different forecast horizons, we consider two summary statistics to evaluate interval-forecast accuracy. The summary statistics chosen are the mean and median values, given by
\begin{align*}
\text{Mean}(\overline{S}_{\alpha}) = \frac{1}{15}\sum^{15}_{h=1}\overline{S}_{\alpha}(h), \qquad
\text{Median}(\overline{S}_{\alpha}) = \overline{S}_{\alpha}[8],
\end{align*}
where $[8]$ represents the $8^{\text{th}}$ term after ranking $\overline{S}_{\alpha}(h)$ for $h=1,2,\dots,15$ from smallest to largest.

\subsection{Comparison of interval-forecast accuracy}\label{sec:interval_compar}

Averaging over all the series at each level of a hierarchy, Table~\ref{tab:HU_LC_interval} presents the mean interval scores $\overline{S}_{\alpha}(h)$ between the Lee--Carter and functional time series methods. The bold entries highlight the method that gives the smallest forecast errors at each level of the hierarchy. Based on the averaged summary statistics of $\overline{S}_{\alpha}(h)$, the functional time series method generally produces more accurate interval forecasts than the ones obtained using the Lee--Carter method. The superiority of the functional time series method is manifested at the bottom level, where the nonparametric smoothing step can assist with modeling and forecasting for those data series that contain a higher level of noise. 

\begin{table}[!h] \centering 
 \caption{Mean interval scores $(\times 100)$ in the holdout sample between the functional time series and Lee--Carter methods applied to the Japanese age-specific mortality rates, using the forecast $S$ matrix. The bold entries highlight the method that gives most accurate summary statistics of the forecasts for each level of the hierarchy and each forecast horizon.} \label{tab:HU_LC_interval} 
  \tabcolsep 0.155in
    \begin{tabular}{@{\extracolsep{5pt}} lcccccc@{}} 
\toprule
 Level & $h=1$ & 5 & 10 & 15 & Mean & Median \\ 
\midrule
\textit{Functional time series method} \\ 
Total & $1.36$ & $2.67$ & $4.31$ & $3.46$ & $\textBF{3.17}$ & $\textBF{3.46}$ \\ 
Sex  & $1.65$ & $3.21$ & $5.41$ & $4.75$ & $\textBF{4.02}$ & $\textBF{4.41}$ \\ 
Region & $2.03$ & $2.98$ & $4.84$ & $4.50$ & $3.82$ & $4.05$ \\ 
Region + Sex & $2.84$ & $3.77$ & $5.71$ & $7.57$ & $4.94$ & $\textBF{5.00}$ \\ 
Prefecture & $3.49$ & $3.95$ & $5.49$ & $7.95$ & $5.06$ & $4.94$ \\ 
Prefecture + Sex  & $5.67$ & $5.90$ & $7.45$ & $10.63$ & $\textBF{7.29}$ & $\textBF{6.82}$ \\ 
\\
\textit{Lee--Carter method} \\
Total & $1.33$ & $2.75$ & $4.87$ & $3.43$ & $3.38$ & $3.52$ \\ 
Sex & $1.64$ & $3.43$ & $6.23$ & $4.28$ & $4.38$ & $\textBF{4.41}$ \\ 
Region & $1.93$ & $2.93$ & $4.94$ & $4.19$ & $\textBF{3.76}$ & $\textBF{4.02}$ \\ 
Region + Sex & $2.69$ & $3.76$ & $6.08$ & $6.09$ & $\textBF{4.90}$ & $5.03$ \\ 
Prefecture & $3.51$ & $4.04$ & $5.52$ & $6.14$ & $\textBF{4.89}$ & $\textBF{4.87}$ \\ 
Prefecture + Sex & $6.08$ & $7.10$ & $9.45$ & $11.52$ & $8.57$ & $8.39$ \\ 
\bottomrule
\end{tabular} 
\end{table} 

Given that the functional time series generally outperforms the Lee--Carter method, we evaluate and compare $\overline{S}_{\alpha}(h)$ among the independent functional time series forecasting, the bottom-up and optimal-combination methods, for each level within the hierarchy of the Japanese data. In Table~\ref{tab:score}, we present the $\overline{S}_{\alpha}(h)$, using the independent functional time series forecasting, bottom-up and optimal-combination methods. The independent forecasting method generally performs the best because it fits each series without the constraint of a hierarchy. At the levels of region and prefecture, the bottom-up method outperforms the independent functional time series forecasting method, which demonstrates the improved interval-forecast accuracy of the bottom-up method when reconciling interval forecasts. Note that the difference between the independent and bottom-up methods is due to randomness in the bootstrapped samples.

\begin{table}[!htbp] \centering 
  \caption{Mean interval scores $(\times 100)$ in the holdout sample among the independent forecasting, bottom-up and optimal-combination methods applied to the Japanese age-specific mortality rates, using the forecast $S$ matrix.}\label{tab:score} 
    \tabcolsep 0.145in
    \begin{tabular}{@{\extracolsep{5pt}} llcccccc@{}} 
\toprule 
Level & Method & $h=1$ & 5 & 10 & 15 & Mean & Median \\ 
\midrule
Total & Base & $1.35$ & $2.65$ & $3.98$ & $4.16$ & $\textBF{3.23}$ & $\textBF{3.64}$ \\ 
& BU & $1.86$ & $2.96$ & $3.94$ & $8.48$ & $3.91$ & $3.72$ \\ 
& OLS & $2.17$ & $3.65$ & $4.68$ & $11.34$ & $4.94$ & $4.61$ \\ \midrule
Sex & Base & $1.63$ & $3.01$ & $4.78$ & $5.58$ & $\textBF{3.84}$ & $\textBF{3.98}$ \\ 
& BU & $1.96$ & $3.44$ & $5.76$ & $12.96$ & $5.42$ & $4.76$ \\ 
& OLS & $2.12$ & $4.10$ & $7.20$ & $15.95$ & $6.71$ & $5.63$ \\ \midrule
Region & Base & $2.01$ & $2.87$ & $4.27$ & $7.63$ & $3.81$ & $3.78$ \\ 
& BU & $2.08$ & $2.59$ & $3.56$ & $9.80$ & $\textBF{3.73}$ & $\textBF{3.24}$ \\ 
& OLS & $2.22$ & $3.20$ & $4.72$ & $12.88$ & $4.84$ & $4.17$ \\ \midrule
Region + Sex & Base & $2.83$ & $3.60$ & $4.89$ & $12.30$ & $\textBF{5.02}$ & $\textBF{4.51}$ \\ 
& BU & $2.83$ & $3.47$ & $5.62$ & $13.89$ & $5.52$ & $4.70$ \\ 
& OLS & $2.72$ & $3.90$ & $6.74$ & $16.96$ & $6.56$ & $5.55$ \\ \midrule
Prefecture & Base & $3.54$ & $3.99$ & $4.90$ & $11.45$ & $5.10$ & $4.57$ \\ 
& BU & $3.52$ & $3.72$ & $4.81$ & $10.13$ & $\textBF{4.87}$ & $\textBF{4.27}$ \\ 
& OLS & $3.21$ & $3.71$ & $4.94$ & $11.68$ & $5.06$ & $4.39$ \\ \midrule
Prefecture + Sex & Base & $5.85$ & $6.18$ & $8.39$ & $16.75$ & $\textBF{8.24}$ & $\textBF{7.30}$ \\ 
& BU & $5.85$ & $6.21$ & $8.42$ & $17.06$ & $8.29$ & $7.33$ \\ 
& OLS & $5.53$ & $6.06$ & $8.37$ & $18.00$ & $8.40$ & $7.32$ \\ 
\bottomrule 
\end{tabular} 
\end{table} 

\subsection{Influence of the $S$ matrix on interval-forecast accuracy}\label{sec:S_interval}

The potential improvement in interval-forecast accuracy in the reconciliation methods relies partially on the accurate forecast of the $S$ matrix. Recall that the $S$ matrix includes ratios of forecast exposure-at-risk. To obtain bootstrapped forecasts of the exposure-at-risk, we use the parametric bootstrap and maximum-entropy bootstrap methods to simulate future samples of the exposure-at-risk on the logarithm scale for age 60. By taking the exponential back-transformation, bootstrapped forecasts of exposure-at-risk on the original scale are obtained. For age above 60, we assume the exposure-to-risk of age $x+1$ in year $t+1$ will be the same as the exposure-to-risk of age $x$ in year $t$. In Table~\ref{tab:actual_S_interval}, we present the mean interval score between the reconciliation methods with the actual holdout $S$ matrix.

\begin{table}[!htbp]
\caption{Mean interval scores $(\times 100)$ in the holdout sample between the bottom-up and optimal-combination methods applied to the Japanese age-specific mortality rates, using the holdout $S$ matrix.}\label{tab:actual_S_interval}
\tabcolsep 0.18in
\centering
\begin{tabular}{@{}llcccccc@{}}
\toprule
Level & Method & $h=1$ & 5 & 10 & 15 & Mean & Median \\
\midrule
Total & BU & 1.85 & 4.45 & 7.44 & 13.40 & 6.56 & 6.47 \\
& OLS & 2.15 & 4.59 & 6.92 & 14.50 & 6.63 & 6.35 \\
\midrule
Sex & BU & 1.96 & 3.66 & 6.10 & 13.32 & 5.68 & 5.08 \\
& OLS & 2.13 & 3.92 & 6.52 & 14.86 & 6.21 & 5.15 \\
\midrule
Region & BU & 2.07 & 3.58 & 6.48 & 14.22 & 5.94 & 5.35 \\
& OLS & 2.21 & 3.92 & 6.64 & 15.73 & 6.29 & 5.59 \\
\midrule
Region + Sex & BU & 2.83 & 3.50 & 5.67 & 13.97 & 5.56 & 4.75 \\
& OLS & 2.72 & 3.65 & 5.93 & 15.73 & 5.96 & 4.99 \\
\midrule
Prefecture & BU & 3.52 & 4.10 & 6.29 & 12.43 & 5.96 & 5.26 \\
& OLS & 3.20 & 4.05 & 6.16 & 13.90 & 5.99 & 5.21 \\
\midrule
Prefecture + Sex & BU & 5.85 & 6.21 & 8.42 & 17.06 & 8.29 & 7.33 \\
& OLS & 5.53 & 5.97 & 8.06 & 17.49 & 8.17 & 7.11\\
\bottomrule
\end{tabular}
\end{table}

We find that the reconciliation methods generally produce more accurate forecasts using the forecast $S$ matrix. This is because the forecast age-specific mortality rates and the $S$ matrix jointly affect the forecast accuracy.

\section{Application to fixed-term annuity pricing}\label{sec:annuity}

An important use of mortality forecasts for elderly (at approximately older than 60 years of age) is in the pension and insurance industries, whose profitability relies on accurate mortality forecasts to appropriately hedge longevity risks. When a person reaches retirement, an optimal way of guaranteeing one individual's financial income in retirement is to purchase an annuity \citep[as demonstrated by][]{Yaari65}. An annuity is a contract offered by insurers guaranteeing a steady stream of payments for either a fixed term or the lifetime of the annuitants in exchange for an initial premium fee. 

In this study, we consider fixed-term annuities, which have grown in popularity in a number of countries, because lifetime immediate annuities, where rates are locked in for life, have been shown to deliver poor value for money \citep[i.e. they may be expensive for the purchaser: see for example][Chapter 6]{CT08}. These fixed-term annuities pay a pre-determined and guaranteed level of income which is higher than the level of income provided by a lifetime annuity. Fixed term annuities provide an alternative to lifetime annuities and allow the purchaser the option of also buying a deferred annuity at a later date. By using the constraint that the terminal age of the fixed term annuity is less than 100, we also avoid the problem of extrapolating the sets of mortality rates up to the very highest ages.

We apply the mortality forecasts to the calculation of a fixed-term annuity \citep[see][p. 114]{DHW09}, and we adopt a cohort approach to the calculation of the survival probabilities. The $\tau$ year survival probability of a person aged $x$ currently at $t=0$ is determined by 
\begin{align*}
_{\tau}p_x &= \prod^{\tau}_{j=1} {}_{1}p_{x+j-1} \\
&= \prod^{\tau}_{j=1} e^{-m_{x+j-1, j-1}},
\end{align*}
which is a random variable given that mortality rates for $j=1,\dots,\tau$ are forecasts obtained by the functional time series method. Here, we assume the central mortality rates are constant throughout the one-year period. 

The price of an annuity with maturity $T$ year, written for an $x$-year-old with benefit \$1 per year and conditional on the path is given by
\begin{align*}
a_{x}^T(\bm{m}_{1:T}^x) &= \sum^{T}_{\tau=1}B(0,\tau)\text{E}(1_{T_x>\tau}|\bm{m}_{1:\tau}^x)  \\
&= \sum^T_{\tau=1}B(0,\tau) {}_{\tau}p_x(\bm{m}_{1:\tau}^x),
\end{align*}
where $B(0,\tau)$ is the $\tau$-year bond price, $\bm{m}_{1:\tau}^x$ is the first $\tau$ elements of $\bm{m}_{1:T}^x$, and ${}_{\tau}p_x(\bm{m}_{1:\tau}^x)$ denotes the survival probability given a random $\bm{m}_{1:\tau}^x$ \citep[see also][]{FPS15}. For the purposes of pricing and risk management, it is vital to produce an accurate forecast of the survival curve ${}_{\tau}p_x$ that best captures the mortality experience of a portfolio.

In Table~\ref{tab:annuity_tab}, to provide an example of the annuity calculations, we compare the best estimate of the annuity prices for different ages and maturities produced by the three forecasting methods for a female policyholder residing in Region 2. We assume a constant interest rate at $\eta = 3\%$ and hence $B(0,\tau) = e^{-\eta\tau}$. Although the annuity price difference might appear to be small, any mispricing can involve a significant risk when considering a large annuity portfolio. Given that an annuity portfolio consists of $N$ policies where the benefit per year is $B$, any underpricing of $\gamma\%$ of the actual annuity price will result in a shortfall of $NBa_x^{T}\gamma/100$, where $a_x^{T}$ is the estimated annuity price being charged with benefit $\$1$ per year. For example, $\gamma=0.1\%$, $N=10,000$ policies written to 85-year-old policyholders with maturity $\tau=15$ years and $\$20,000$ benefit per year will result in a shortfall of $10,000 \times 20,000 \times  8.3649 \times 0.1\% = 1.67$ million.

\begin{center}
  \tabcolsep 0.23in
\begin{longtable}{@{\extracolsep{5pt}} lcccccc@{}} 
  \caption{Estimates of annuity prices with different ages and maturities ($T$) for a female policyholder residing in Region 2. These estimates are based on forecast mortality rates from 2015 to 2055. We consider only contracts with maturity so that age + maturity $\leq 100$. If age + maturity $> 100$, NA will be shown in the table.} 
  \label{tab:annuity_tab} 
\\[-1.8ex]
\toprule \\[-1.8ex] 
& $T=5$ & $T=10$ & $T=15$ & $T=20$ & $T=25$ & $T=30$ \\\midrule
\endfirsthead
& $T=5$ & $T=10$ & $T=15$ & $T=20$ & $T=25$ & $T=30$ \\
\endhead
\hline \multicolumn{7}{r}{{Continued on next page}} 
\endfoot
\endlastfoot
& \multicolumn{6}{c}{age = 60} \\\midrule
Base & 4.5249 & 8.3364 & 11.5272 & 14.1688 & 16.2995 & 17.9099 \\ 
BU & 4.5256 & 8.3380 & 11.5284 & 14.1659 & 16.2878 & 17.8890 \\ 
OLS & 4.5274 & 8.3452 & 11.5461 & 14.1999 & 16.3465 & 17.9824 \\
\midrule
& \multicolumn{6}{c}{age = 65} \\\midrule
Base & 4.5125 & 8.2903 & 11.4178 & 13.9404 & 15.8470 & 17.0969 \\
BU & 4.5127 & 8.2892 & 11.4111 & 13.9227 & 15.8181 & 17.0693 \\ 
OLS & 4.5146 & 8.2998 & 11.4381 & 13.9765 & 15.9111 & 17.2066 \\
\midrule
& \multicolumn{6}{c}{age = 70} \\\midrule
Base & 4.4946 & 8.2156 & 11.2169 & 13.4854 & 14.9724 & 15.7251 \\ 
BU & 4.4930 & 8.2073 & 11.1955 & 13.4504 & 14.9391 & 15.7133 \\ 
OLS & 4.4983 & 8.2278 & 11.2444 & 13.5434 & 15.0830 & 15.8936 \\ 
\midrule 
& \multicolumn{6}{c}{age = 75} \\\midrule
Base & 4.4596 & 8.0567 & 10.7754 & 12.5577 & 13.4597 & NA \\
BU & 4.4551 & 8.0392 & 10.7438 & 12.5294 & 13.4579 & NA \\
OLS & 4.4632 & 8.0733 & 10.8247 & 12.6672 & 13.6372 & NA \\
\midrule
& \multicolumn{6}{c}{age = 80} \\\midrule
Base & 4.3769 & 7.6850 & 9.8537 & 10.9513 & NA & NA \\
BU & 4.3694 & 7.6667 & 9.8435 & 10.9755 & NA & NA \\
OLS & 4.3849 & 7.7268 & 9.9647 & 11.1430 & NA & NA \\
 \midrule
& \multicolumn{6}{c}{age = 85} \\\midrule
Base & 4.1765 & 6.9144 & 8.3001 & NA & NA & NA \\
BU & 4.1752 & 6.9315 & 8.3649 & NA & NA & NA \\
OLS &  4.2011 & 7.0144 & 8.4956 & NA & NA & NA \\
\midrule
& \multicolumn{6}{c}{age = 90} \\\midrule
Base & 3.7912 & 5.7100 & NA & NA & NA & NA \\
BU & 3.8142 & 5.7977 & NA & NA & NA & NA \\
OLS & 3.8449 & 5.8692 & NA & NA & NA & NA \\
\midrule
& \multicolumn{6}{c}{age = 95} \\\midrule
Base & 3.2044 & NA & NA & NA & NA & NA \\ 
BU & 3.2686 & NA & NA & NA & NA & NA \\ 
OLS & 3.2847 & NA & NA & NA & NA & NA \\\bottomrule
\end{longtable}
\end{center}

To measure forecast uncertainty, we construct the bootstrapped prediction intervals of age-specific mortality rates, derive the survival probabilities and calculate the corresponding annuities associated with different ages and maturities. Given that we have only 40 years of data, we construct one-step-ahead to 15-step-ahead bootstrapped forecasts of age-specific mortality rates. In Table~\ref{tab:annuitiy_CI}, we present the pointwise 95\% prediction intervals of annuities for different ages and maturities, where age + maturity $\leq$ 75. Although the difference in prediction intervals of annuity prices among the three methods might appear to be small, as stated, any mispricing can have a dramatic impact on the forecast uncertainty of actual annuities when considering a large annuity portfolio.

\begin{table}[!t]
\centering
  \tabcolsep 0.42in
  \caption{95\% pointwise prediction intervals of annuity prices with different ages and maturities ($T$) for female policyholder residing in Region 2, for example. These estimates are based on forecast mortality rates from 2015 to 2029. We only consider contracts with maturity so that age + maturity $\leq 75$. If age + maturity $> 75$, NA will be shown in the table.}\label{tab:annuitiy_CI}
\begin{tabular}{@{\extracolsep{5pt}} lccc@{}} 
\toprule
& $T=5$ & $T=10$ & $T=15$  \\\midrule
& \multicolumn{3}{c}{age = 60} \\\midrule
Base & (4.5184, 4.5327) & (8.3241, 8.3604) & (11.5048, 11.5770) \\
BU & (4.5228, 4.5317) & (8.3320, 8.3569) & (11.5213, 11.5643) \\
OLS & (4.5236, 4.5313) & (8.3360, 8.3564) & (11.5300, 11.5658) \\
\midrule
& \multicolumn{3}{c}{age = 65} \\\midrule
Base & (4.5024, 4.5254) & (8.2634, 8.3287) & NA \\
BU & (4.5076, 4.5211) & (8.2765, 8.3138) & NA \\
OLS & (4.5097, 4.5208) & (8.2849, 8.3159) & NA \\
\midrule
& \multicolumn{3}{c}{age = 70} \\\midrule
Base & (4.4762, 4.5160) & NA & NA \\
BU & (4.4859, 4.5049) & NA & NA \\
OLS & (4.4897, 4.5075) & NA & NA \\
\bottomrule
\end{tabular}
\end{table}

\section{Conclusions}\label{sec:conclu}

Using the national and subnational Japanese mortality data, we evaluate and compare the point-forecast accuracy between the Lee--Carter and the functional time series methods. Based on the forecast-accuracy criteria, we find that the functional time series method outperforms the Lee--Carter method. The superiority of the functional time series method is driven by the use of nonparametric smoothing techniques to deal with noisy mortality rates at older ages, in particular for males; and more than one component is used to achieve improved model fitting.

By using the functional time series method to produce base forecasts, we consider the issue of forecast reconciliation by applying two grouped functional time series forecasting methods, namely the bottom-up and optimal-combination methods. The bottom-up method models and forecasts data series at the most disaggregated level, and then aggregates the forecasts using the summing matrix constructed on the basis of forecast exposure-to-risk. The optimal-combination method combines the base forecasts obtained from independent functional time series forecasting methods using linear regression. The optimal-combination method generates a set of revised forecasts that are as close as possible to the base forecasts, but that also aggregates consistently with the known grouping structure. Under some mild assumptions, the regression coefficient can be estimated by ordinary least-squares.

Using the Japanese data, we compare the one-step-ahead to 15-step-ahead forecast accuracy between the independent functional time series forecasting method and two proposed grouped functional time series forecasting methods. We find that the grouped functional time series forecasting methods produce more accurate point forecasts than those obtained by the independent functional time series forecasting method, averaged over all levels of the hierarchy. In addition, the grouped functional time series forecasting methods produce forecasts that obey the natural group structure, thus giving forecast mortality at the subnational levels that add up to the forecast mortality rates at the national level. Between the two grouped functional time series forecasting methods, the bottom-up method is recommended for the data we considered. 

We apply the independent functional time series and two grouped functional time series methods to forecast age-specific mortality rates from 2015 to 2055. We then calculate the cumulative survival probability and obtain the fixed-term annuity prices. As expected, we find that the cumulative survival probability has a pronounced impact on annuity prices. Although annuity prices do not differ greatly for the mortality forecasts obtained by the three methods, mispricing could have a dramatic impact for a portfolio, particularly when the yearly benefit is a great deal larger than $\$1$ dollar per year. To assess forecast uncertainty, we obtain bootstrapped forecasts of age-specific mortality rates, derive their survival probabilities and calculate their annuity prices. By taking quantiles, the pointwise prediction intervals of annuity prices can be constructed for various ages and maturities. 

There are several ways in which this paper can be further extended, and we briefly discuss three here. First, the Lee--Carter and functional time series models are not new-data invariant, as demonstrated in \cite{CLL14}. When an additional year of mortality data becomes available and models are updated, the principal component scores in the previous years will be affected. The time-varying nature of the principal component scores generally provides better goodness-of-fit and forecast accuracy compared to the model that satisfies the new-data-invariant property. When the focus is on the tractability of the resulting mortality indexes, future research should implement \citeauthor{CBD06}'s \citeyearpar{CBD06} original model to forecast mortality rates. Second, subject to the availability of data, the hierarchy can be disaggregated more finely by considering different causes of death \citep{ML97, GS15} or socioeconomic status \citep{BBA02, SAS+13, VH14}. Third, the methodology can be applied to calculate other types of annuity prices, such as the whole-life immediate annuity or deferred annuity.

\section*{Acknowledgments}

The authors are grateful for the insightful comments and suggestions of the two reviewers, as well as the seminar participants at the Cass Business School at the City University of London, the Economic and Social Research Council Centre for Population Change at the University of Southampton, the School of Mathematical Sciences at the University of Adelaide, and the Department of Applied Finance and Actuarial Studies at the Macquarie University. The research is funded by a research school grant from the Australian National University. 

\bigskip
\begin{center}
{\large\bf SUPPLEMENTARY MATERIAL}
\end{center}

\begin{description}

\item[Results for the sensitivity analysis of retained number of components]  The results of point and interval forecasts produced by different values of $J$. (sensitivity.pdf)

\item[R-package for functional time series forecasting] The R-package \textit{ftsa} containing code to produce point forecasts from the Lee--Carter and functional time series forecasting methods described in the article. The R-package can be obtained from the Comprehensive R Archive Network (\url{https://cran.r-project.org/web/packages/ftsa/index.html}).

\item[Code for grouped functional time series forecasting] The R code to produce point forecasts from the two grouped functional time series forecasts described in the article. (gfts.zip)

\item[Code for Shiny application] The R code to produce a Shiny user interface for plotting every series in the Japanese data hierarchy. (shiny.zip)

\end{description}

\newpage
\bibliographystyle{agsm}
\bibliography{GFTS}

\end{document}